\documentclass[useAMS,usenatbib,twocolumn]{mn2e}
\usepackage{amsmath}
\usepackage{amssymb}
\usepackage{graphicx}
\usepackage{color}

\newcommand{\bfu}{\mathbf{u}}
\newcommand{\bfB}{\mathbf{B}}
\newcommand{\Ma}{\mathcal{M}}
\newcommand{\cs}{c_\mathrm{s}}
\newcommand{\R}{\mathcal{R}}
\newcommand{\Rb}{\left|\mathcal{R}\right|}

\title[]{Statistical analysis of the mass-to-flux ratio in turbulent cores: effects of magnetic field reversals and dynamo amplification}
\author[Bertram et al.]{E.~Bertram$^{1}$, C.~Federrath$^{1,2,3}$, R.~Banerjee$^{1,4}$ \& R.~S.~Klessen$^{1}$\\
$^1$Zentrum f\"ur Astronomie der Universit\"at Heidelberg, Institut f\"ur Theoretische  Astrophysik, Albert-Ueberle-Str.~2, 69120 Heidelberg, Germany\\
$^2$Ecole Normale Sup\'{e}rieure de Lyon, CRAL, 69364 Lyon, France\\
$^3$Monash Centre for Astrophysics (MoCA), School of Mathematical Sciences, Monash University, Vic 3800, Australia\\
$^4$Hamburger Sternwarte, Gojenbergsweg 112, 21029 Hamburg, Germany
}

\begin{document}

\maketitle

\abstract
We study the mass-to-flux ratio $M/\Phi$ of clumps and cores in simulations of supersonic, magnetohydrodynamical turbulence for different initial magnetic field strengths. We investigate whether the $M/\Phi$-ratio of core and envelope, $\R = (M/\Phi)_{core}/(M/\Phi)_{envelope}$ can be used to distinguish between theories of ambipolar diffusion and turbulence-regulated star formation. We analyse $\R$ for different Lines-of-Sight (LoS) in various sub-cubes of our simulation box. We find that, 1) the average and median values of $\Rb$ for different times and initial magnetic field strengths are typically $\gtrsim1$, 2) the average and median values of $\Rb$ saturate at $\overline{\Rb} \approx 1$ for smaller magnetic fields, 3) values of $\Rb<1$ for small magnetic fields in the envelope are caused by field reversals when turbulence twists the field lines such that field components in different directions average out. Finally, we propose two mechanisms for generating values $\Rb\lesssim1$ for the weak and strong magnetic field limit in the context of a turbulent model. First, in the weak field limit, the small-scale turbulent dynamo leads to a significantly increased flux in the core and we find $\Rb\lesssim1$. Second, in the strong field limit, field reversals in the envelope also lead to values $\Rb\lesssim1$. These reversals are less likely to occur in the core region where the velocity field is more coherent and the internal velocity dispersion is typically subsonic.
\endabstract

\section{Introduction}

Understanding star formation is a fundamental problem for theoretical astrophysics~\citep[see reviews by][]{MacLowAndKlessen,McKeeAndOstriker}. For several years, the idea was that star formation is mainly regulated by the magnetic field and ambipolar diffusion~\citep[e.g.][]{Mouschovias1984,ShuAdamsLizano1987}. In this model, the neutral particles, not directly affected by the magnetic field in a gas cloud, slowly move inwards to the gravitational centre, while a significant part of the magnetic flux remains in the envelope of the collapsing cloud. Alternatively, star formation could be regulated by supersonic turbulence~\citep[][]{MacLowAndKlessen}. \citet{PadoanAndNordlund} and \citet{PadoanEtAl} developed a super-Alfv\'{e}nic, turbulent model of dark clouds. Those turbulence-regulated models of star formation predict that clumps and cores form through turbulent compression at the intersection of shocks \citep{ballesteros03,klessen05}.

Both models, ambipolar diffusion and turbulence try to explain the relatively low star formation rate \citep{ZuckermanAndEvans1974} observed in the Galaxy, however, the physical processes put forward in the two models are fundamentally different. The main problem is that the amount of turbulent, kinetic energy and the amount of magnetic energy is typically observed to be of the same order of magnitude in interstellar clouds \citep{Crutcher1999}. Thus, it is still an open question to which extend ambipolar diffusion and supersonic turbulence regulate star formation.

In order to test the two star formation models described above, \citet{Crutcher} introduced a new quantity, $\R = (M/\Phi)_{\rm{core}}/(M/\Phi)_{\rm{envelope}}$, which is the mass-to-flux ratio of the core and envelope of a dense clump. If ambipolar diffusion plays the central role in the process of star formation, we would expect a value of $\Rb>1$. This is because the clouds are initially supported by the magnetic field, but then ambipolar diffusion finally leads to an increase of the mass relative to the flux in the centre (core) of the cloud. Thus, a significant fraction of the magnetic flux remains in the envelope during the contraction, and consequently $\Rb>1$. In contrast, if the clumps form in a super-Alfv\'{e}nic, turbulent medium, we would expect values of $\Rb\lesssim1$, i.e., a mass-to-flux ratio on average being higher in the envelope than in the core~\citep{Lunttila,Crutcher}. This is attributed to field reversals in the envelope of the cloud, due to the larger amounts of turbulence there, compared to the denser core. There are three reasons to expect $\Rb\lesssim1$. First, the envelopes of the cores typically have a much higher turbulent velocity dispersion than the interior \citep{BensonAndMyers1989,AndreEtAt2007,LadaEtAl2008,BeutherAndHenning2009,SmithEtAl2009,FederrathEtAl2010}. Second, the magnetic field in the core is stronger than in the envelope due to the compression of the field lines. Both aspects make field reversals more likely in the envelope of the cloud, such that magnetic field lines can cancel out there, leading to $\Rb\lesssim1$. The third important mechanism leading to $\Rb\lesssim1$ is small-scale dynamo action \citep[e.g.,][]{Brandenburg} in the core. This process leads to an efficient amplification of the magnetic flux in the core, relative to the envelope as shown in the simulations by \citet{SurEtAl2010} and \citet{FederrathEtAl2011}, or investigated analytically by \citet{schleicher10}.

The idea of this paper is to perform supersonic, magnetohydrodynamical (MHD) simulations of turbulence in dense clouds, and to analyse the relative mass-to-flux ratios in the core and envelope, $\R$, of clumps identified in the simulations. We use a large statistical sample in PPP (Position-Position-Position) and PP (Position-Position) space to obtain statistically significant results. We take into account different initial magnetic field strengths (i.e., weak and strong magnetic fields) and investigate the dependence of our results on the resolution of the simulations. In addition, we compare two different methods for computing $\R$, in order to test the validity of our conclusions. Finally, we compare our results to those given by \citet{Crutcher} and \citet{Lunttila}, and discuss the implications of our results on $\R$ as a measure to distinguish between ambipolar diffusion and turbulence-regulated star formation.

In section~\ref{sec:methods}, we describe the MHD simulations, the clump finding and analysis, and introduce two different methods for computing $\R$. In section~\ref{sec:results}, we show that our turbulence simulations produce clumps with mean values of $\Rb\gtrsim1$ for runs with relatively strong initial magnetic fields. Runs with weak magnetic fields produce a larger fraction of cores with $\Rb\lesssim1$, but the mean value is still slightly higher than unity for both analysis methods of $\Rb$. In section~\ref{sec:discussion} and~\ref{sec:summary}, we present our conclusions and describe two possible mechanisms, field reversals and the small-scale dynamo, for generating values $\Rb\lesssim1$.

\section{Methods}
\label{sec:methods}

In the following, we describe our numerical methods used to model supersonic, magnetohydrodynamical turbulence and the analysis performed to define cores and magnetic field strengths along the Line-of-Sight (LoS), and to compute mass-to-flux ratios in the cores and envelopes found in the simulations.

\subsection{MHD simulations of driven, supersonic turbulence}

We computed numerical solutions of the compressible, three-dimensional, ideal magnetohydrodynamical equations with the grid code FLASH v2.5 \citep{FryxellEtAl2000}, here written in a form where the permeability constant $\mu_0 = 1$:
\begin{equation}
\def\arraystretch{1.2}
\begin{array}{@{}l@{}}
\partial_t \rho + \nabla\cdot\left(\rho \bfu\right)=0 ,\\

\partial_t\!\left(\rho \bfu\right) + \nabla\cdot\left(\rho \bfu\!\otimes\!\bfu - \bfB\!\otimes\!\bfB\right) + \nabla p_\star =  \rho{\bf F}, \\

\partial_t E + \nabla\cdot\left[\left(E+p_\star\right)\bfu - \left(\bfB\cdot\bfu\right)\bfB\right] = 0, \\

\partial_t \bfB + \nabla\cdot\left(\bfu\!\otimes\!\bfB - \bfB\!\otimes\!\bfu\right) = 0, \\
\nabla\cdot\bfB = 0,
\end{array}
\label{eq:mhd}
\end{equation}
where $\rho$, $\bfu$, $p_\star=p+ (1/2)\left|\bfB\right|^2$, $\bfB$, and $E=\rho \epsilon_\mathrm{int} + (1/2)\rho\left|\bfu\right|^2 + (1/2)\left|\bfB\right|^2$ denote density, velocity, pressure (thermal plus magnetic), magnetic field, and total energy density (internal, kinetic, and magnetic), respectively. The MHD equations were closed with a polytropic equation of state, $p=\cs^2\rho$, such that the gas remains isothermal with constant sound speed $\cs=0.2\,\mathrm{km}\,\mathrm{s}^{-1}$, assuming a constant gas temperature of $11.2\,\mathrm{K}$. To drive turbulence, we apply the forcing term ${\bf F}$ as a source term in the momentum equation above. The forcing term is modelled with a stochastic Ornstein-Uhlenbeck process \citep{EswaranPope1988,SchmidtEtAl2009,FederrathEtAl2009}, such that ${\bf F}$ varies smoothly in space and time with an autocorrelation time equal to the eddy-turnover time, $T=L/(2\Ma\cs)$ at the largest scales, $L/2$, in the periodic simulation domain of size $L=4\,\mathrm{pc}$. $\Ma=u_\mathrm{rms}/\cs$ denotes the root-mean-squared (rms) Mach number, the ratio of the rms velocity and the sound speed. All models were driven to an rms Mach number of $\Ma\approx10$, typical for interstellar clouds \citep[e.g.,][]{CsengeriEtAl}. Turbulence is fully developed at $t=2\,T$ \citep{SchmidtEtAl2009,FederrathEtAl2010,PriceFederrath2010}. We thus analyse our results in this statistically steady regime for $t=2.0$, 2.4, and $2.8\,T$ to explore the temporal variations of our results.

The turbulent forcing is constructed in Fourier space such that kinetic energy is only injected at the smallest wave numbers, $1<\left|\mathbf{k}\right|L/2\pi<3$, i.e., the largest scales. Construction in Fourier space allows us to decompose the force field into its solenoidal (rotational) and compressible (dilatational) parts. In this study, however, we only use solenoidal (divergence-free) forcing of the turbulence and leave the study of the mass-to-flux ratios in compressively driven MHD turbulence for future work. 

We used the new HLL3R scheme for ideal MHD, developed by \citet{Waagan2009}, and tested extensively in the FLASH code by \citet{WaaganFederrathKlingenberg2011}. The scheme makes use of a novel approximate Riemann solver for ideal MHD \citep{Bouchut1,Bouchut2} that preserves positive states in highly supersonic MHD turbulence. To explore the influence of different initial magnetic field strengths, we performed simulations with an initial plasma $\beta_0=2p_0/B_0^2=0.01$, 0.1, 1, 10, and 100, corresponding to initial field strengths of $B_0=44$, 14, 4.4, 1.4, and $0.44\,\mu\mathrm{G}$. All models were evolved on a fixed grid with $256^3$ grid zones. For the model with $\beta_0=1$ we additionally performed a resolution study with $128^3$, $256^3$, and $512^3$ grid cells in Appendix~\ref{app:resol}.

Figure~\ref{fig:betaAlfvenMach_t} shows the time evolution of the plasma beta and the Alfv\'enic Mach number. The rms sonic Mach number is not shown, but settles around $\Ma\approx10$ within the first two turbulent turnover times, $t\gtrsim2\,T$. After that, the turbulence is fully developed, and thus we restrict our analysis to $t=2.0$, 2.4, and $2.8\,T$ to explore the temporal variations. Figure~\ref{fig:betaAlfvenMach_t} shows that all models are super-Alfv\'enic, except for the run with $\beta_0=0.01$, which approaches an Alfv\'enic Mach number of $\Ma_\mathcal{A}\approx0.8$.

\begin{figure}
\begin{center}
\includegraphics[width=1.0\linewidth]{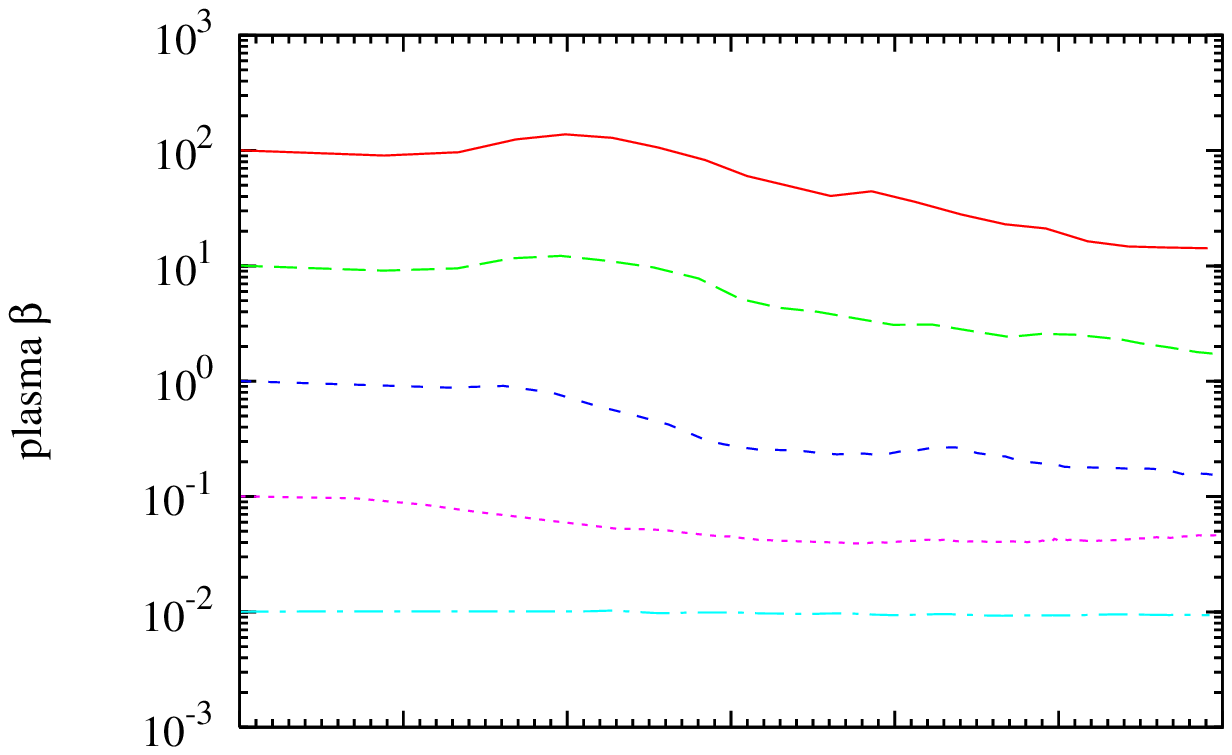} \\
\includegraphics[width=1.0\linewidth]{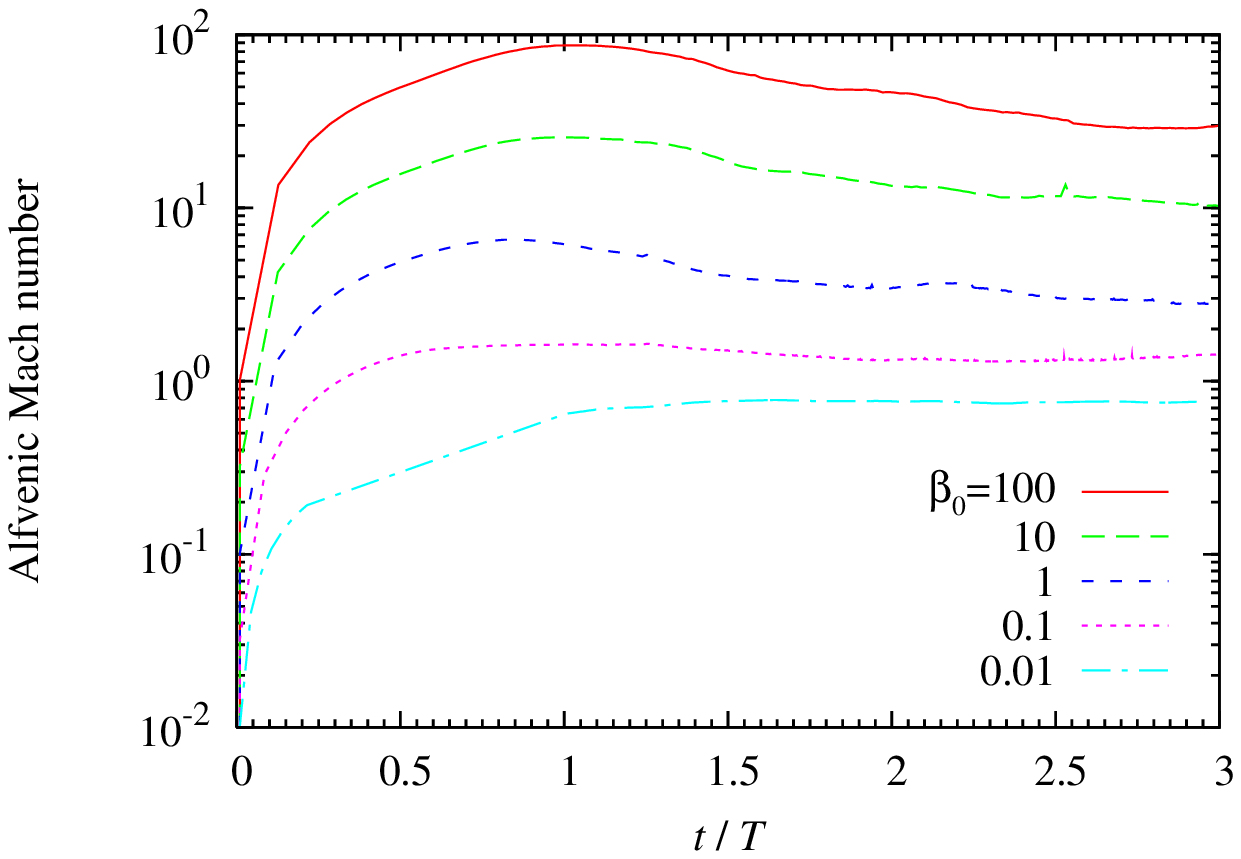}
\end{center}
\caption{Top: Evolution of the ratio of thermal to magnetic pressure, plasma $\beta$, for all MHD models with different initial values, $\beta_0$. Bottom: Same as top panel, but for the Alfv\'enic Mach number.}
\label{fig:betaAlfvenMach_t}
\end{figure}

\subsection{Core analysis of the numerical MHD simulations}
\label{subsec:coreanalysis}

In the following we will use the definition of a clump and a core as it was given by \citet{Bergin}. They define a clump as an object with a size of $0.3$--$3\,\mathrm{pc}$ and a core as an object with a size of $0.03$--$0.2\,\mathrm{pc}$. This means, that in every simulation, we will extract clumps that contain cores which will be used for further computations.

We consider the full 3-dimensional information in the Position-Position-Position (PPP) as well as the corresponding surface density maps, i.e., the 2-dimensional projection of Position-Position (PP), and we note that only the latter is accessible to the observations. For the PPP case in our simulation box with $256^3$ grid cells, we look for the densest cell and select a small cube with an edge length of $15$ grid cells around it (the density maximum is centred in the middle of this cube). The only information that we need for our calculations are the density and the magnetic field components. For our resolution studies with $128^3$ and $512^3$ grid cells, the small cubes are also scaled with a factor of 2 in each spatial direction, so that we have $7^3$ and $29^3$ grid cells for each clump/core, respectively. The small cube is extracted out of the simulation box such that it leaves an empty region in our box that will not be used any further. Then we find the next dense region and extract another clump and so on. In total, we select as many clumps out of each simulation box as we can find at any given time and initial magnetic field strength by their density maximum for the PPP measurements, under the assumption that the peak density does not fall below $\rho = 20 \, \overline \rho$, where $\overline \rho$ is the mean density in the computational domain. We are able to extract about 100 clumps out of each simulation snapshot for the PPP case. For the PP measurements we choose a column density threshold of $\Sigma = 2 \overline \Sigma$, where $\overline \Sigma$ is the mean surface density in the simulation. With this threshold, we are able to extract about 40 clumps out of each simulation. The mean (column-) density of the clumps always ranges between $10^{-19}$ and $10^{-20}\,\mathrm{g\,cm^{-3}}$ for the PPP case and between 0.1 and $1.2\,\mathrm{g\,cm^{-2}}$ for the PP case.

Our clump selection algorithm is somewhat arbitrary. The selection of overdensities in a complex, filamentary structure, typical of molecular clouds, is however a general problem \citep{Smith,PinedaEtAl2009,SchmidtEtAl2010}. In order to test at least one other way of defining our clumps, we use clumpfind, a 'friend-of-friend' algorithm \citep[e.g.,][]{Williams,Klessen}, and perform the same analysis. Clumpfind extracts clumps that are identified as connected regions. We locate the density maximum in each clump and define a grid cell being part of a core, if the density is larger than two thirds of the maximum density in the clump. In analogy, the envelope is defined as all grid cells with density below one third of the maximum density. In this way, we test different core-to-envelope volume ratios. Using clumpfind affects the particular selection of clumps and hence their individual properties. However, we find that the overall statistical properties of the clump ensemble are not significantly affected by using the clumpfind selection algorithm, and hence our main conclusions remain intact.

Following \citet{Crutcher}, we define the quantity $\R$, which compares the mass-to-flux ratio of core and envelope of our clump,
\begin{equation}
\label{eq:M2F}
\R =  \frac{[M/\Phi]_{core}}{[M/\Phi]_{envelope}} = \frac{[\Sigma/B_{LoS}]_{core}}{[\Sigma/B_{LoS}]_{envelope}}
\end{equation}
where the column density $\Sigma$ denotes the Line-of-Sight (LoS) integral over the density, $\Sigma = \int_{LoS} \rho dz$, which is computed along any LoS of our homogeneous grid and $B_{LoS}$ is the magnetic field component in the LoS-direction in which we observe our clumps. For a more realistic treatment of this scenario, all magnetic field components in each LoS are mass-weighted with the density of each grid cell, to make our computations more comparable to observed Zeeman-splitting measurements:
\begin{equation}
B_{LoS} = \frac{1}{\Sigma} \int_{LoS} \rho B_{z} dz\;.
\end{equation}
To calculate $\R$, we first have to define a region for the core and the envelope, respectively. Our aim in general is to apply the method given by \citet{Crutcher}, who used one telescope beam for the core and four somewhat larger telescope beams for the envelope.

\begin{figure}
\centerline{
\includegraphics[width=0.7\linewidth]{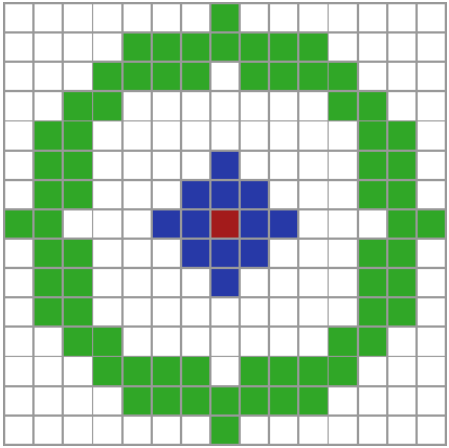}
}
\caption{Definition of core and envelope in our clumps for a resolution of $256^3$ grid cells. The density peak is located in the centre, the blue cells define our core size, while the green cells define our envelope.}
\label{fig:clump_structure}
\end{figure}

Since we have many more beams (grid cells) available in the simulations, we use them to increase the statistical significance of the measurement. Figure~\ref{fig:clump_structure} shows how we defined core and envelope in our clumps. After contraction (integration) of each 3D-cube along the Line-of-Sight, we choose a circle core size of 5 grid cells in diameter around the density maximum for the $256^3$ simulation (3 cells for the $128^3$ simulation and 11 cells for the $512^3$ simulation) and neglect the complex geometry of the cloud. The envelope, which is treated as a thin shell instead of four telescope beams as done by \citet{Crutcher}, has a size of 2 grid cells (1 cell and 5 cells). These values reflect the average sizes of cores and envelopes in our simulations. Our results do not depend significantly on this choice of grid cells of core and envelope. Table~\ref{tab:core_prop} gives a short overview of some mean properties of clumps for a resolution of $256^3$ grid cells for the PPP and PP case. An alternative method for computing $\R$ will be described in the next section.

\begin{table}
\begin{tabular}{|l|l|l|}
\hline\hline
 & PPP & PP \\
\hline
 Cells in diameter & 5 & 5\\
 Physical Diameter & $0.08\,\mathrm{pc}$ & $0.08\,\mathrm{pc}$\\
 Mass & $0.3$--$3\,$M$_{\sun}$ & $2$--$29\,$M$_{\sun}$\\
 Density & $10^{-19}$--$10^{-20}\,\mathrm{g\,cm^{-3}}$ & $0.1$--$1.2\,\mathrm{g\,cm^{-2}}$\\
\hline\hline
\end{tabular}
\caption{Mean properties of our selected clumps for a resolution of $256^3$ grid cells for the PPP and PP case.}
\label{tab:core_prop}
\end{table}

\subsection{Two different methods of computing $\R$}
\label{subsec:computeR}

In general, $\R$ is a statistical quantity and there are several possibilities to calculate its value \citep[see, e.g., the discussion about $M/\Phi$ in][]{Vazquez2011}. Here we adopt and compare two different methods. First, we compute the average of the magnetic field components in the LoS in the core and envelope, $B_{LoS} = \frac{1}{N}\sum B_i$, and afterwards calculate $\R$ by using eq.~(\ref{eq:M2F}). Second, we obtain $\R$ pixelwise, i.e., we select one pixel of our core and one pixel of the envelope and compute $\R_i$ only for those two pixels. We do that for all $M$ possible combinations of pixels of core and envelope and take the logarithmic average of all absolute values afterwards, because we obtain a very wide range of $\R$-values, which are rather logarithmically than linearly distributed. We do the same for the magnetic field and also compute the logarithmic average for $N$ absolute field components in the envelope:
\begin{equation}
\begin{array}{@{}l@{}}
\R = 10^{\frac{1}{M}\sum_i \log{|\R_i}|} \\
B_{LoS} = 10^{\frac{1}{N}\sum_i \log{|B_i|}}
\end{array}
\end{equation}
Our aim is to compare the results of both methods of computing $\R$ and $B_{LoS}$ with each other.

\subsection{Computation of field reversals}

For our first analysis method, we also define the following quantity, $X(N)$, which tells us how many field reversals appear along the Line-Of-Sight (LoS):
\begin{equation}
\label{fieldreversals}
X(N) = \frac{N_{+} - N_{-}}{N_{+} + N_{-}}
\end{equation}
Here $N_{+}$ is the total number of cells with a positive sign of the magnetic field component in LoS in the envelope and $N_{-}$ the total number of cells with a negative sign of the magnetic field component in LoS in the envelope. $N = N_{+} + N_{-}$ is the total number of grid cells (the normalisation) counted in the envelope. By definition, $X$ can only lie in a range between -1 to +1, corresponding to either cells only with a negative or positive magnetic field component. A value of $X$ close to zero indicates that there are many field reversals in the envelope, i.e., nearly as much cells with a positive and a negative sign. In the following, we only consider the absolute values of $\Rb$ and $|B_{LoS}|$. However, we emphasise that both quantities are computed with their individual signs (per beam and per cell) taken into account. This is important, because of possible cancellations of the $B$-field along the LoS due to field reversals.

\begin{figure*}
\centerline{
\includegraphics[width=0.5\linewidth]{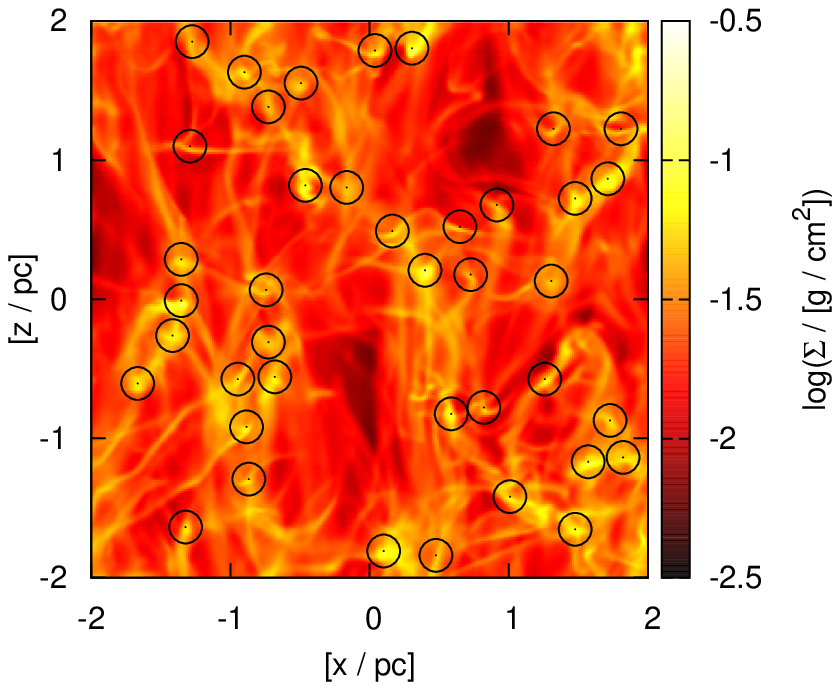}
\includegraphics[width=0.5\linewidth]{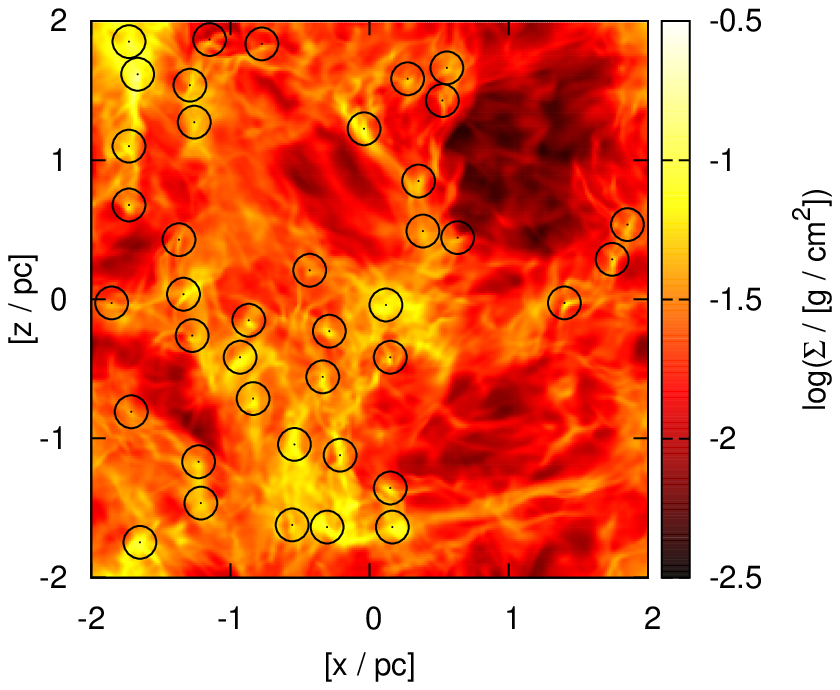}
}
\caption{Logarithmic column density map computed along the LoS in y-direction for an initial plasma beta of $\beta_0 = 0.01$ (i.e. for a very strong field, left), and $\beta_0 = 100$ (i.e. for a very weak field, right) at $t = 2.0\,T$. For $\beta_0 = 0.01$, one can see the outstanding z-direction of the magnetic field, while for $\beta_0 = 100$, the magnetic field is so weak that turbulence can easily tangle the magnetic field lines, such that the overall density structure is rather isotropic. Also labelled are the positions of 40 density-peaks (dot in the middle of each circle), which fulfil our threshold condition and the maximum diameter of the envelopes (circles).}
\label{fig:CD_Sim_0.01}
\end{figure*}

\section{Results}
\label{sec:results}

\subsection{PP column density map and large scale structures}

As an example, Figure~\ref{fig:CD_Sim_0.01} shows a logarithmic column density map of our simulations with $\beta_0 = 0.01$ (left) and $\beta_0 = 100$ (right), i.e., for a very strong and very weak initial magnetic field. The left plot shows clear filamentary structures that are oriented preferably along the z-direction in which our initial field was established. For comparison, the right plot shows a rather isotropic structure, which is caused by weak magnetic field lines that can easily be tangled by turbulence. Also labelled are the positions of the density peaks and the maximum diameter of the envelopes of our extracted clumps for the PP case.

\subsection{Time evolution}

When comparing different simulation snapshots, we do not find any systematic time-dependence in the distribution of our clumps for PPP and PP for any $\beta_0$. To illustrate this point, Table~\ref{tab:values2} shows the statistical core values for different times ($t=2.0, 2.4, 2.8\,T$) for two extreme initial values of $\beta_0 = 0.01$ and $\beta_0 = 100$ in the PPP case, computed with the first analysis method. On the basis of a $1\sigma$-error interval, no significant change in the distribution of our clumps (except for the usual statistical fluctuations) with respect to time are found.

\subsection{Scaling of the average values of $B_{LoS}$ and $\R$}

The plots in Figure~\ref{fig:scatter1} and~\ref{fig:scatter2} are computed with our first analysis method (see sec.~\ref{subsec:computeR}). Figure~\ref{fig:scatter1} shows a clear trend between the value of $\Rb$ and $|B_{LoS}|$ in each direction, both for PPP and PP. The absolute values of the magnetic field are smaller for smaller absolute values of $\R$. For a small initial $\beta_0 = 0.01$ (first row) one has very strong magnetic field lines in the simulation in z-direction that lead to a very compact distribution of clumps for every time. Here the magnetic field lines are so strong that turbulence is not able to stir the medium perpendicular to that direction. To quantify this effect, we show the number of field reversals $X$ defined in eq.~(\ref{fieldreversals}) in Figure~\ref{fig:scatter2}, such that we can finally compare which clumps in Figure~\ref{fig:scatter1} have a certain number of field reversals or not. As expected, we do not find any field reversals for the clumps with very low $\beta_0$, observed in z-direction, which means that $X = 1$ (first row, Figure~\ref{fig:scatter2}). All clumps observed in the other directions are isotropically distributed. This is because of our initial direction of the magnetic field in z-direction at the beginning of our simulation. If we increase the initial value of plasma $\beta$, we find that the average value of the magnetic field in LoS of our clumps decreases. In Figure~\ref{fig:scatter1}, we also added the observational results of the four clouds from \citet{Crutcher}, L1448CO, B217-2, L1544 and B1, which fit into our general trend of increasing $\R$ with increasing $B_{LoS}$. However, the observed values are at the lower end of our distribution. We do not find any significant differences between the PPP and PP measurements. We also varied the number of cells of core and envelope, as described in section~\ref{subsec:coreanalysis}, but could not find any significant change in the distribution of the clumps.

We might expect that the average values of the mean magnetic field of the clumps shown in Figure~\ref{fig:scatter1} should scale like $\overline B_{LoS,i}/\overline B_{LoS,j} = \sqrt{\beta_{0,j} / \beta_{0,i}}$ because $\beta \propto B^{-2}$, where $i$ and $j$ denote simulations with different initial plasma $\beta$. Table~\ref{tab:values1} gives an overview of the average values of the magnetic field, $\R$, and their standard deviations and medians. If we consider the z-direction, we should always obtain a constant ratio of $\overline B_{LoS,i} / \overline B_{LoS,j} = \sqrt{10} \approx 3.2$, if $i$ and $j$ correspond to $\beta_0 = 0.01$ and 0.1, 0.1 and 1 and so on. Therefore, the ratio from $i=0.01$ to $j=0.1$ is $\overline B_{LoS,0.01} / \overline B_{LoS,0.1} = 46.2 / 15.8 \approx 3$, which fits well to our theoretically predicted value of 3.2. For the other ratios, we get values of 1.7, 1.7 and 1.8 respectively. This discrepancy comes from the fact that the magnetic field is amplified by the small-scale dynamo \citep{Brandenburg} in cases of high initial $\beta_0$, i.e., $\beta$ is a function of time, as is the Alfv\'enic Mach number (see Figure~\ref{fig:betaAlfvenMach_t}).

\begin{table}
\begin{tabular}{|l|l||l|l|l||l|l||l}
\hline\hline
Time & LoS & $\overline {|B_{LoS}|}$ & $|\tilde{B}_{LoS}|$ & $\sigma_{|B|}$ & $\overline {\Rb}$ & $\tilde {\Rb}$ & $\sigma_{\Rb}$ \\
\hline
$\beta_0 = 0.01$ & & & & & & & \\
\hline
 & x & 10.3 & 9.3 & 7.5 & 4.0 & 2.2 & 6.4\\
 2.0 & y & 7.1 & 5.7 & 5.6 & 3.8 & 2.6 & 3.9\\
 & z & 46.2 & 46.0 & 7.6 & 2.8 & 2.7 & 1.1\\
\hline
 & x & 7.5 & 5.8 & 5.9 & 4.2 & 2.2 & 6.3\\
 2.4 & y & 8.4 & 7.8 & 6.3 & 4.9 & 2.5 & 8.3\\
 & z & 46.0 & 46.0 & 8.1 & 2.8 & 2.7 & 1.2\\
\hline
 & x & 8.5 & 8.0 & 6.2 & 4.4 & 2.0 & 9.5\\
 2.8 & y & 8.1 & 7.0 & 6.5 & 4.3 & 2.6 & 9.6\\
 & z & 46.0 & 46.4 & 9.5 & 2.7 & 2.5 & 1.1\\
\hline
$\beta_0 = 100$ & & & & & & & \\
\hline
 & x & 2.7 & 1.9 & 2.4 & 1.3 & 0.8 & 2.0\\
 2.0 & y & 3.2 & 2.2 & 3.0 & 2.5 & 1.0 & 7.2\\
 & z & 3.1 & 2.1 & 3.4 & 1.9 & 0.9 & 3.6\\
\hline
 & x & 3.4 & 2.8 & 2.9 & 1.1 & 0.9 & 1.0\\
 2.4 & y & 3.4 & 2.4 & 3.2 & 2.4 & 1.0 & 6.4\\
 & z & 3.0 & 2.7 & 2.4 & 1.7 & 1.0 & 2.5\\
\hline
 & x & 2.5 & 1.9 & 2.2 & 1.9 & 0.8 & 4.3\\
 2.8 & y & 3.0 & 2.4 & 2.5 & 2.5 & 1.0 & 5.8\\
 & z & 2.6 & 2.2 & 2.2 & 2.1 & 0.8 & 6.3\\
\hline\hline
\end{tabular}
\caption{Mean, median and standard deviation for all directions of the magnetic field component and $\Rb$ for $256^3$ cells for $\beta_0 = 0.01$ and $\beta_0 = 100$ for the PPP case, computed with our first analysis method described in section~\ref{sec:methods}. From top to bottom (separated by a line): values for different times, $t=2.0, 2.4, 2.8\,T$. All values of $B$ are given in $\mu$G and time in $T$.}
\label{tab:values2}
\end{table}

\begin{table}
\begin{tabular}{|l|l||l|l|l||l|l||l}
\hline\hline
$\beta_0$ & LoS & $\overline {|B_{LoS}|}$ & $|\tilde{B}_{LoS}|$ & $\sigma_{|B|}$ & $\overline {\Rb}$ & $\tilde {\Rb}$ & $\sigma_{\Rb}$ \\
\hline
 & x & 10.3 & 9.3 & 7.5 & 4.0 & 2.2 & 6.4\\
 0.01 & y & 7.1 & 5.7 & 5.6 & 3.8 & 2.6 & 3.9\\
 & z & 46.2 & 46.0 & 7.6 & 2.8 & 2.7 & 1.1\\
\hline
 & x & 10.6 & 9.6 & 7.5 & 3.4 & 2.1 & 4.4\\
 0.1 & y & 9.3 & 7.5 & 7.5 & 3.6 & 2.1 & 5.1\\
 & z & 15.8 & 15.7 & 8.2 & 3.1 & 2.2 & 3.1\\
\hline
 & x & 7.6 & 5.4 & 6.9 & 3.4 & 1.5 & 6.0\\
 1 & y & 9.3 & 8.1 & 6.6 & 3.7 & 1.9 & 9.0\\
 & z & 9.5 & 8.5 & 6.5 & 3.7 & 1.9 & 8.0\\
\hline
 & x & 4.9 & 3.6 & 4.8 & 2.5 & 1.1 & 5.9\\
 10 & y & 6.5 & 4.8 & 6.3 & 3.5 & 1.4 & 9.1\\
 & z & 5.6 & 4.3 & 4.7 & 1.7 & 1.4 & 1.8\\
\hline
 & x & 2.7 & 1.9 & 2.4 & 1.3 & 0.8 & 2.0\\
 100 & y & 3.2 & 2.2 & 3.0 & 2.5 & 1.0 & 7.2\\
 & z & 3.1 & 2.1 & 3.4 & 1.9 & 0.9 & 3.6\\
\hline\hline
\end{tabular}
\caption{Mean, median and standard deviation for all directions for the absolute magnetic field component and $\Rb$ for $256^3$ cells and for $t = 2.0\,T$ for the PPP case, computed with our first analysis method described in section~\ref{sec:methods}. From top to bottom (separated by a line): values for an initial plasma $\beta$ of 0.01, 0.1, 1, 10 and 100. All values of $B$ are given in $\mu$G.}
\label{tab:values1}
\end{table}

\subsection{Effect of field reversals on $\R$}
\label{sec:fieldreversals}

\begin{figure*}
\centerline{
\includegraphics[height=0.22\linewidth]{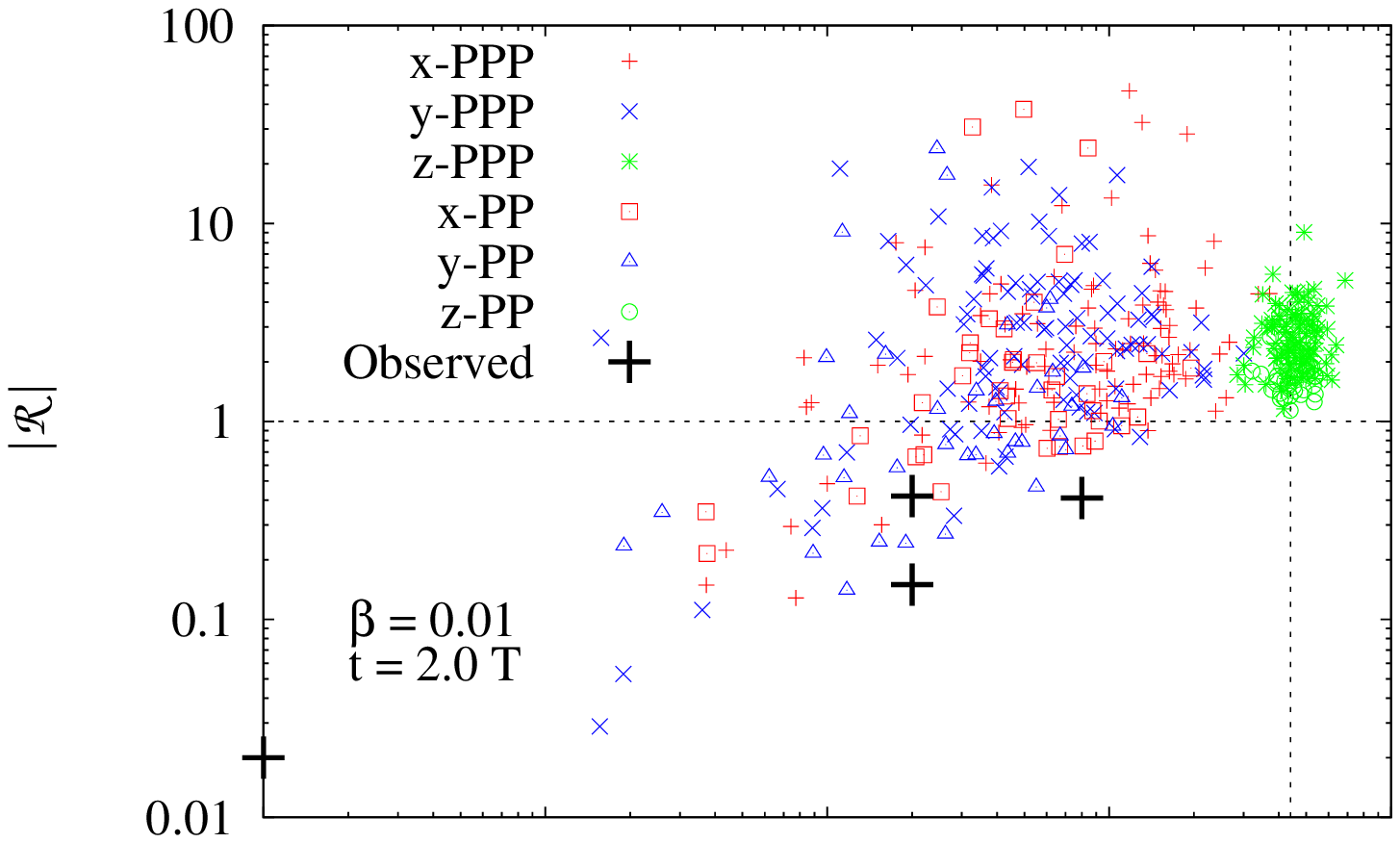}
\includegraphics[height=0.22\linewidth]{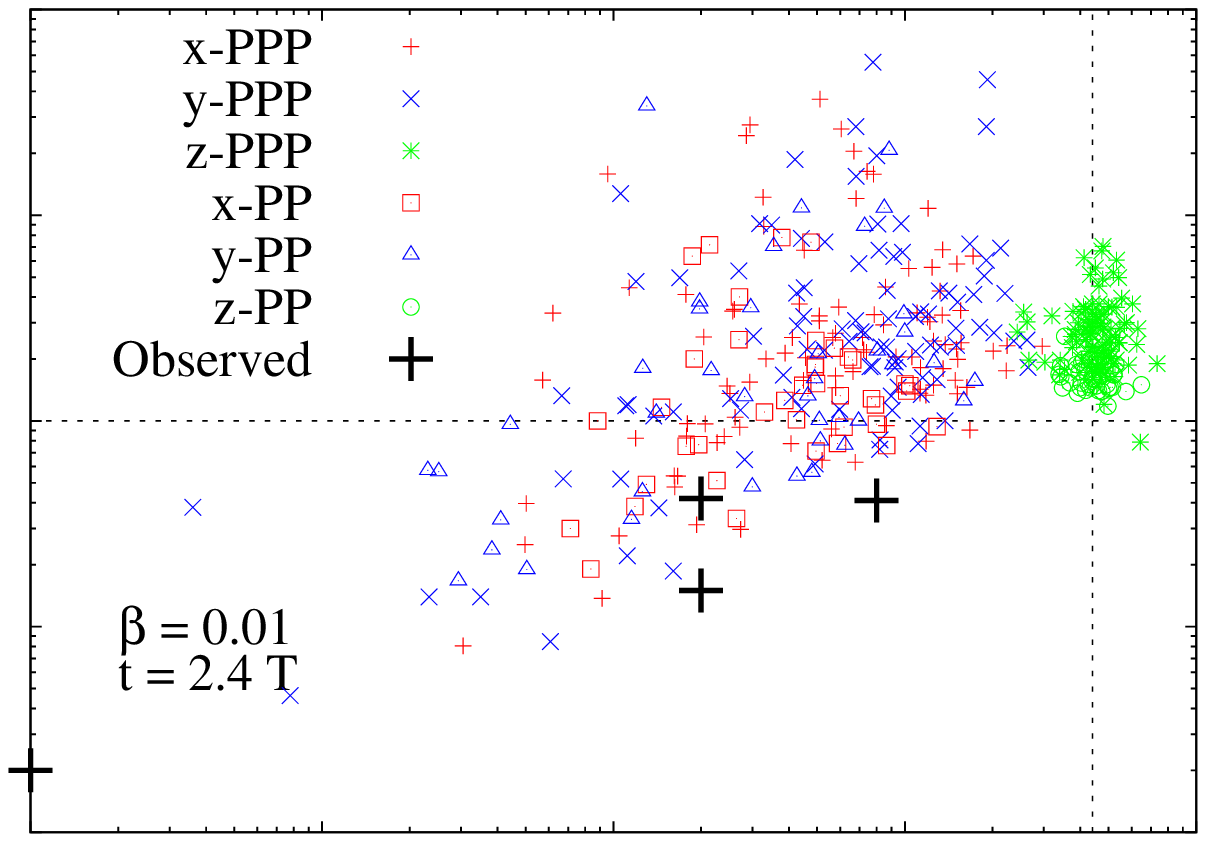}
\includegraphics[height=0.22\linewidth]{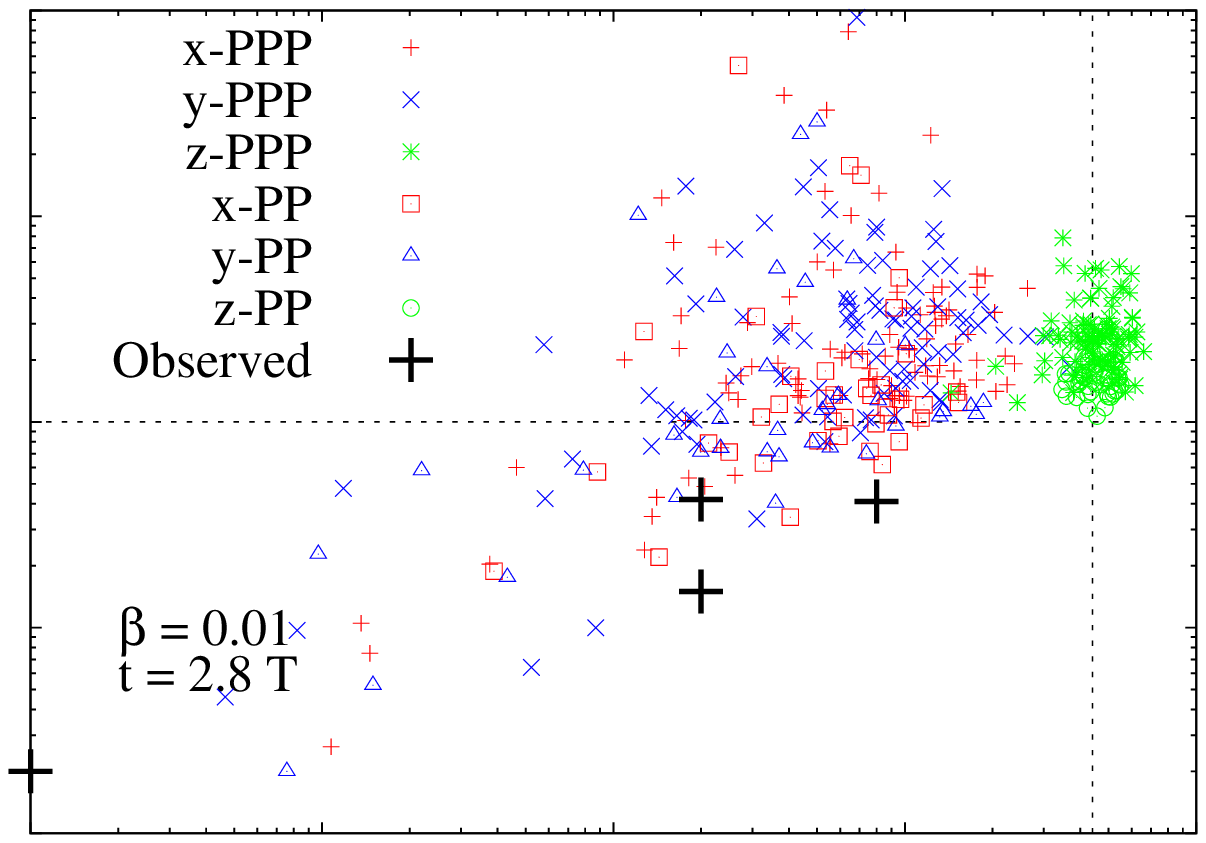}
}

\centerline{
\includegraphics[height=0.22\linewidth]{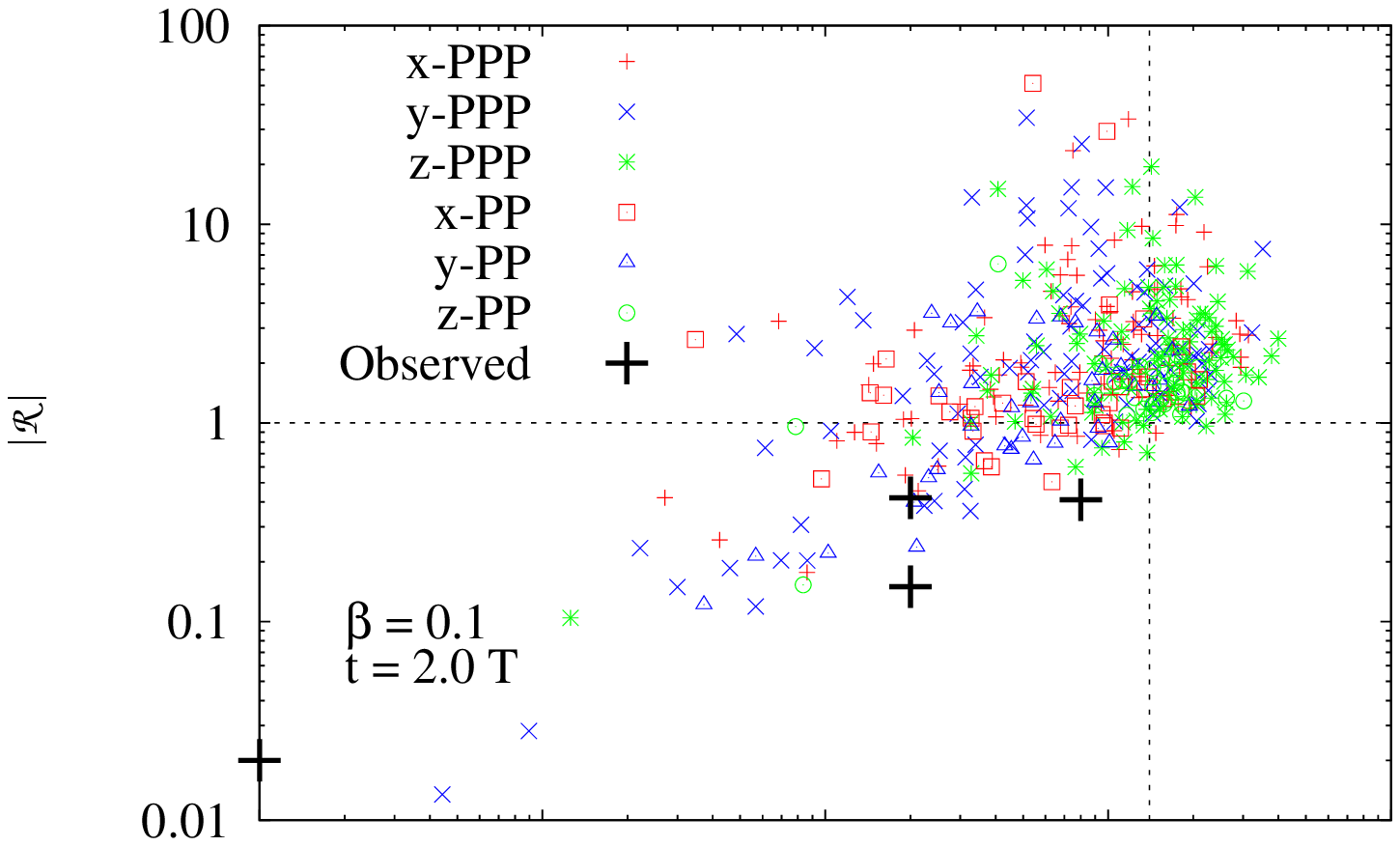}
\includegraphics[height=0.22\linewidth]{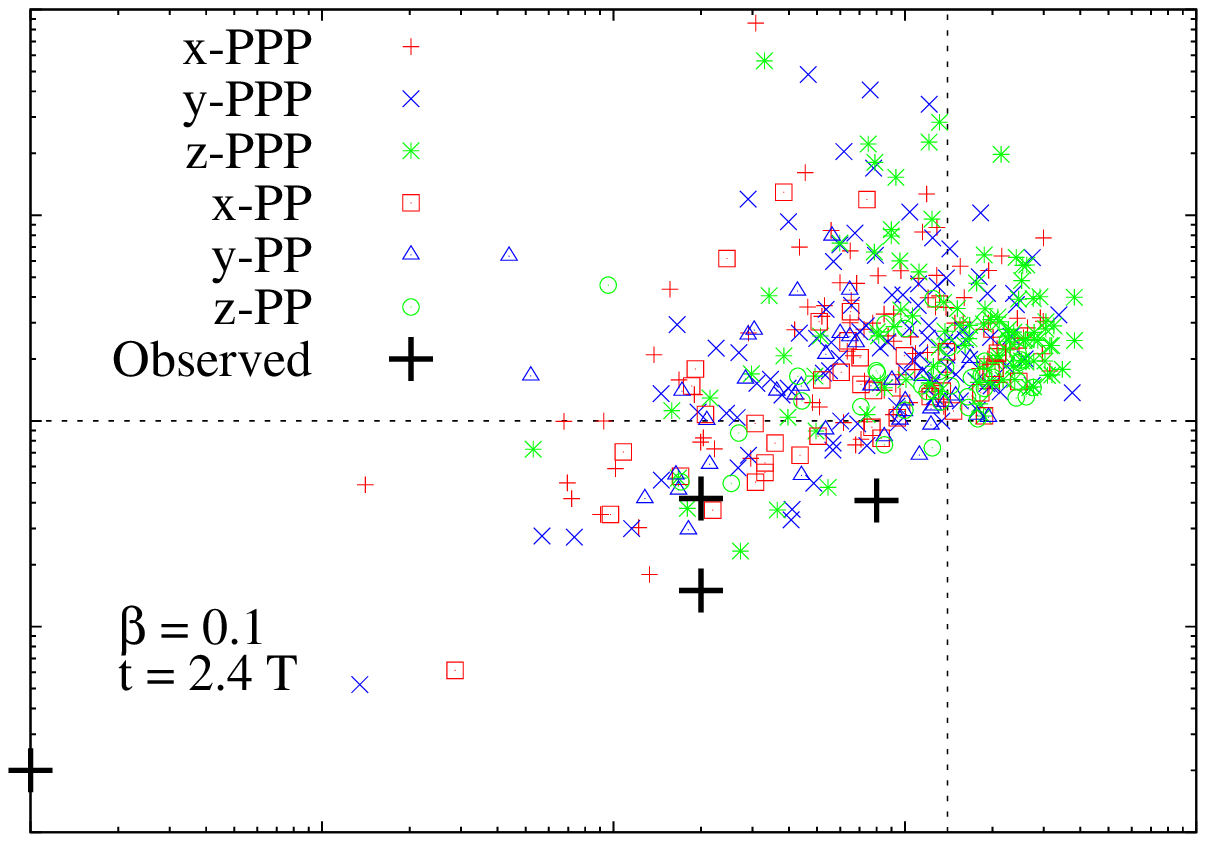}
\includegraphics[height=0.22\linewidth]{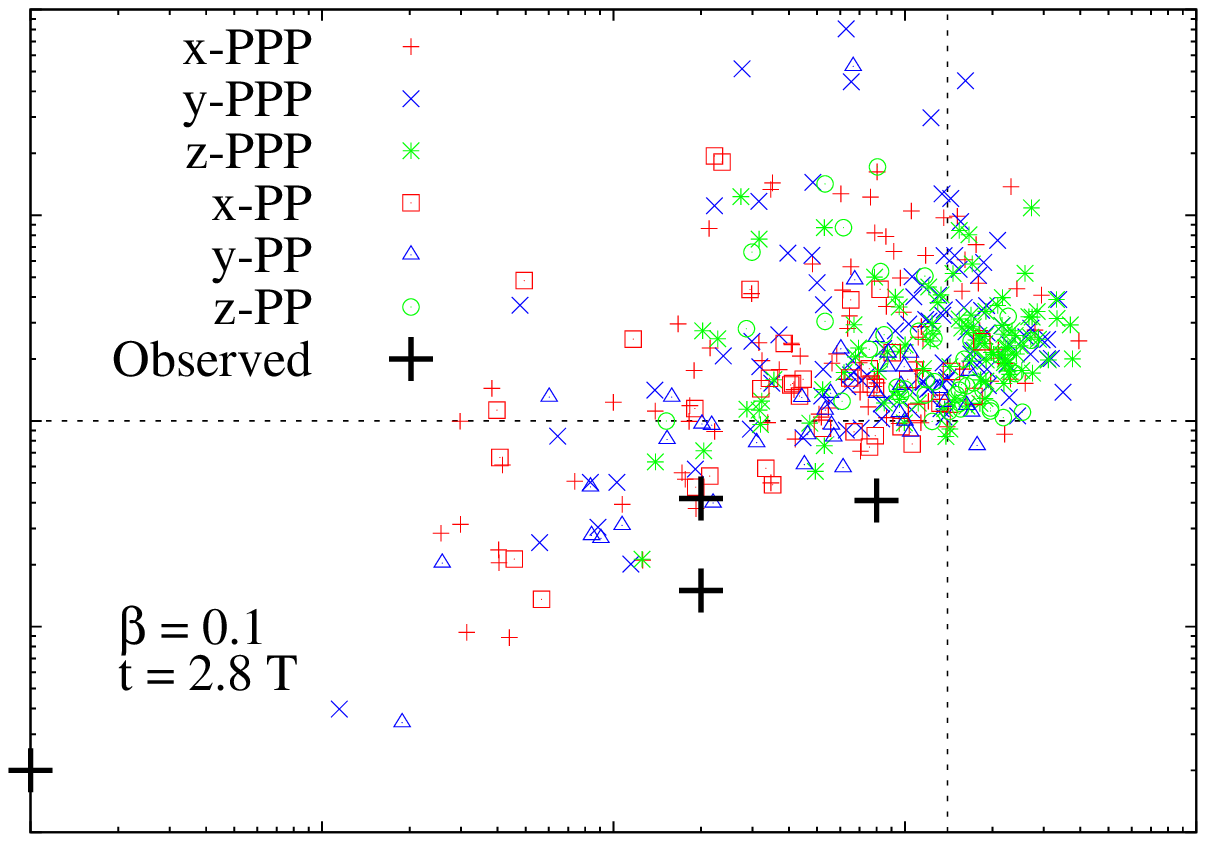}
}

\centerline{
\includegraphics[height=0.22\linewidth]{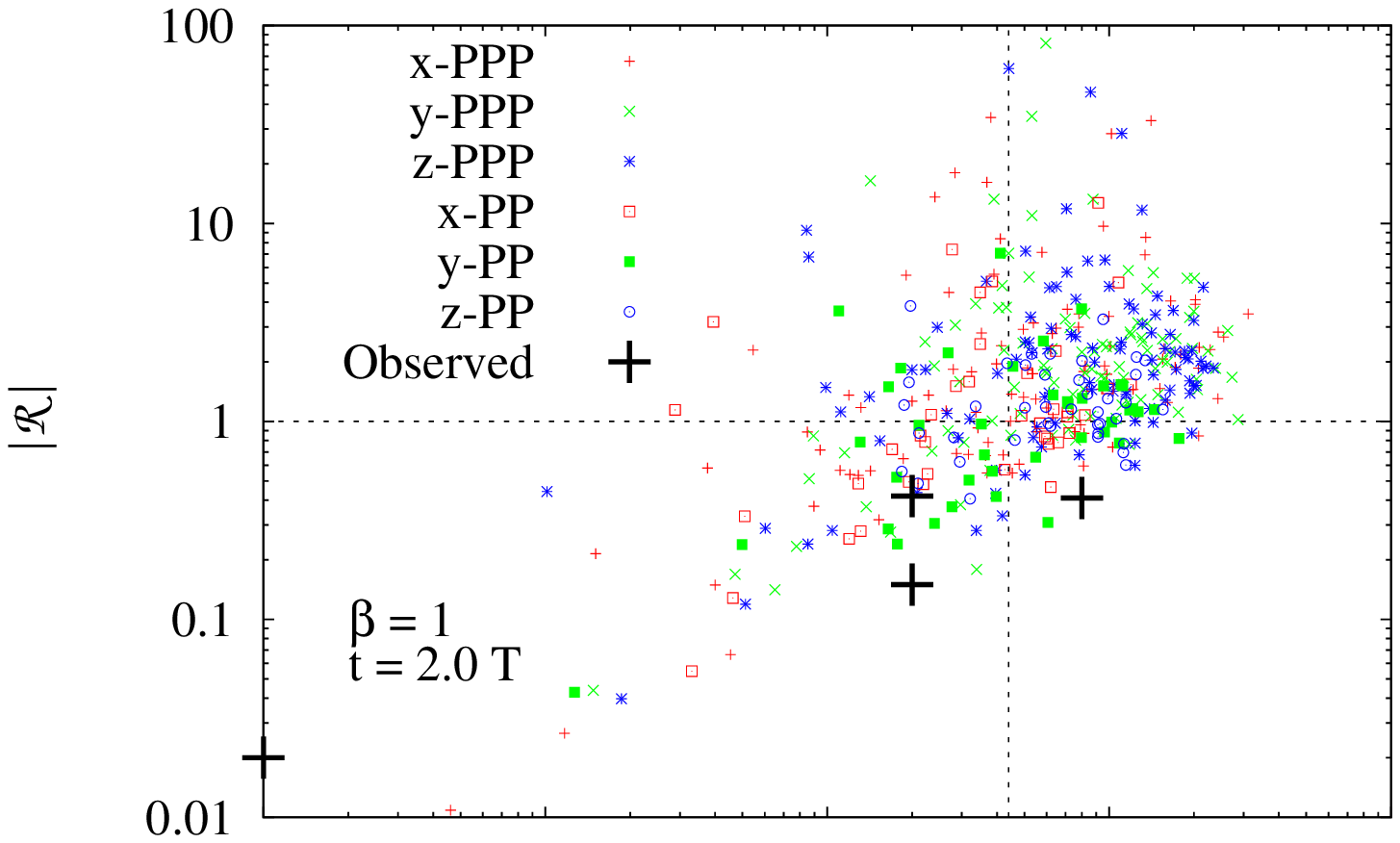}
\includegraphics[height=0.22\linewidth]{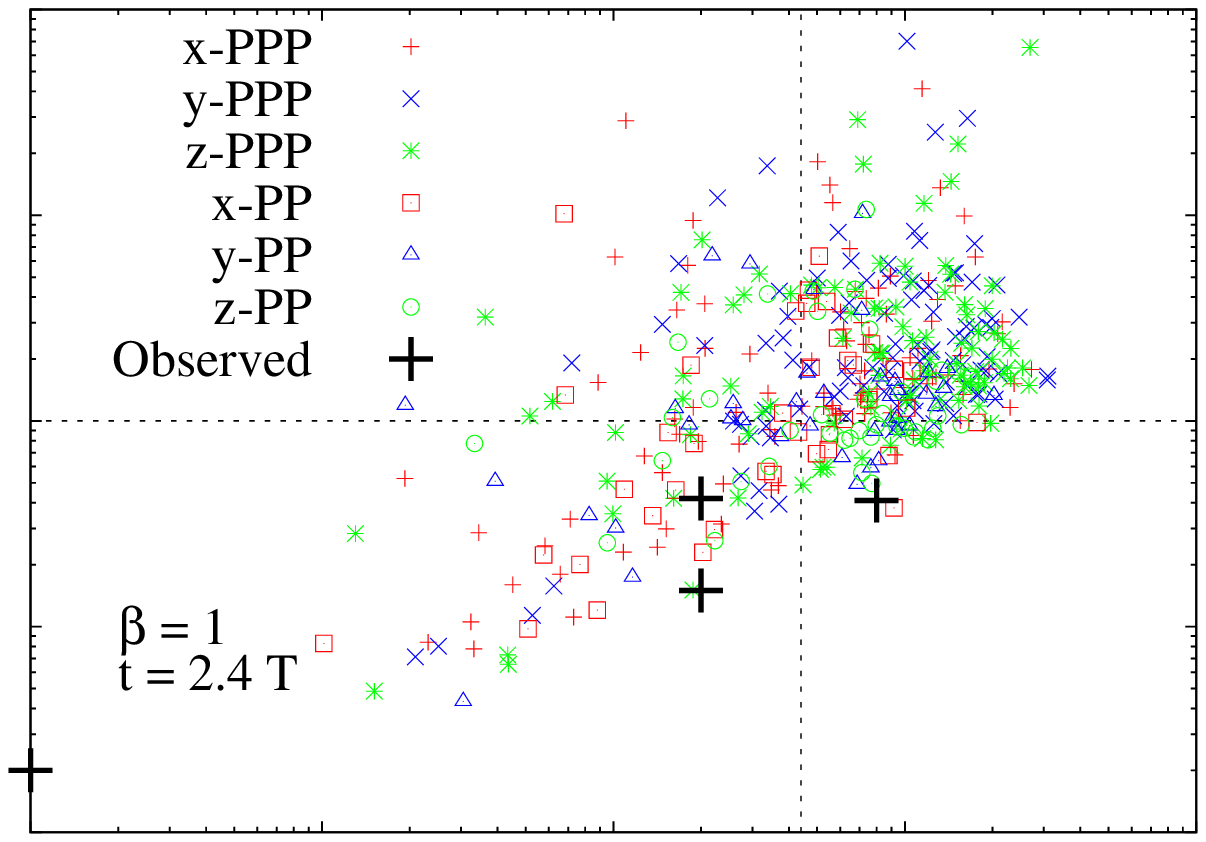}
\includegraphics[height=0.22\linewidth]{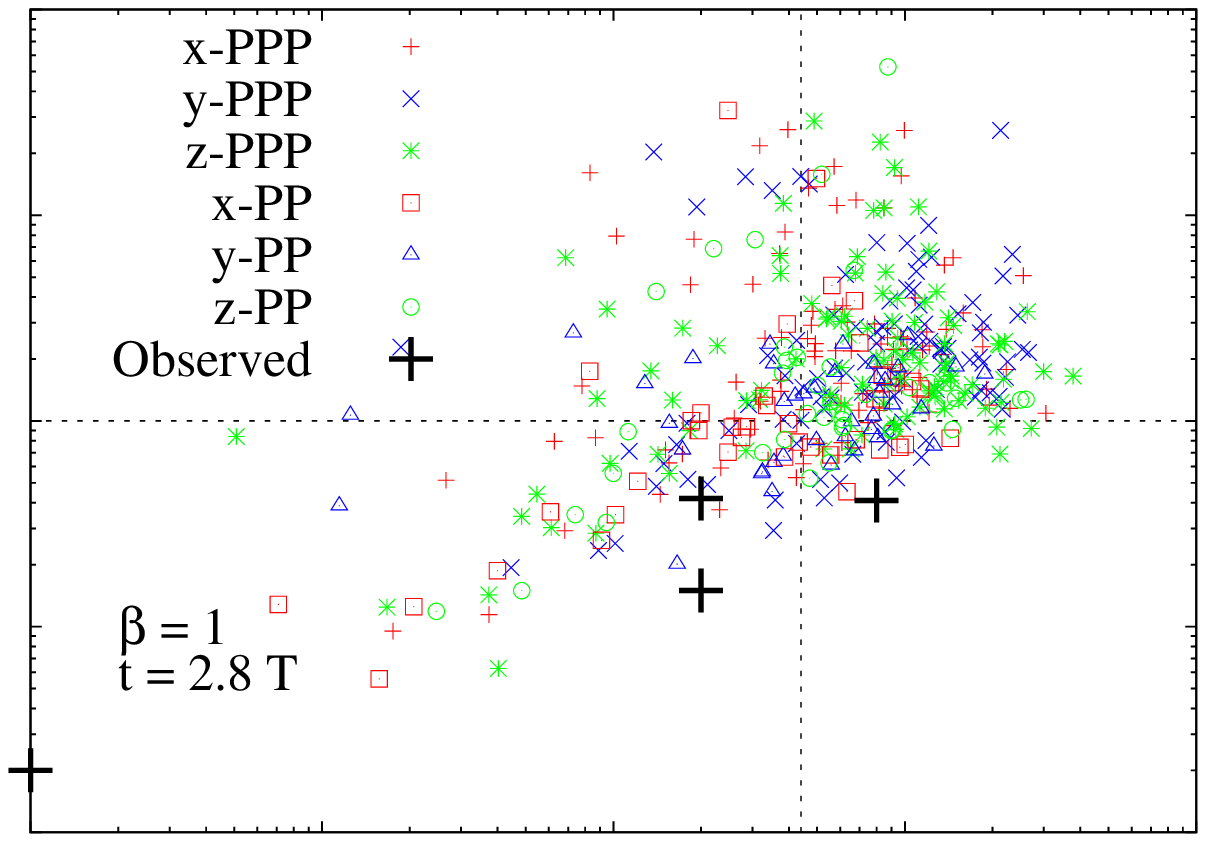}
}

\centerline{
\includegraphics[height=0.22\linewidth]{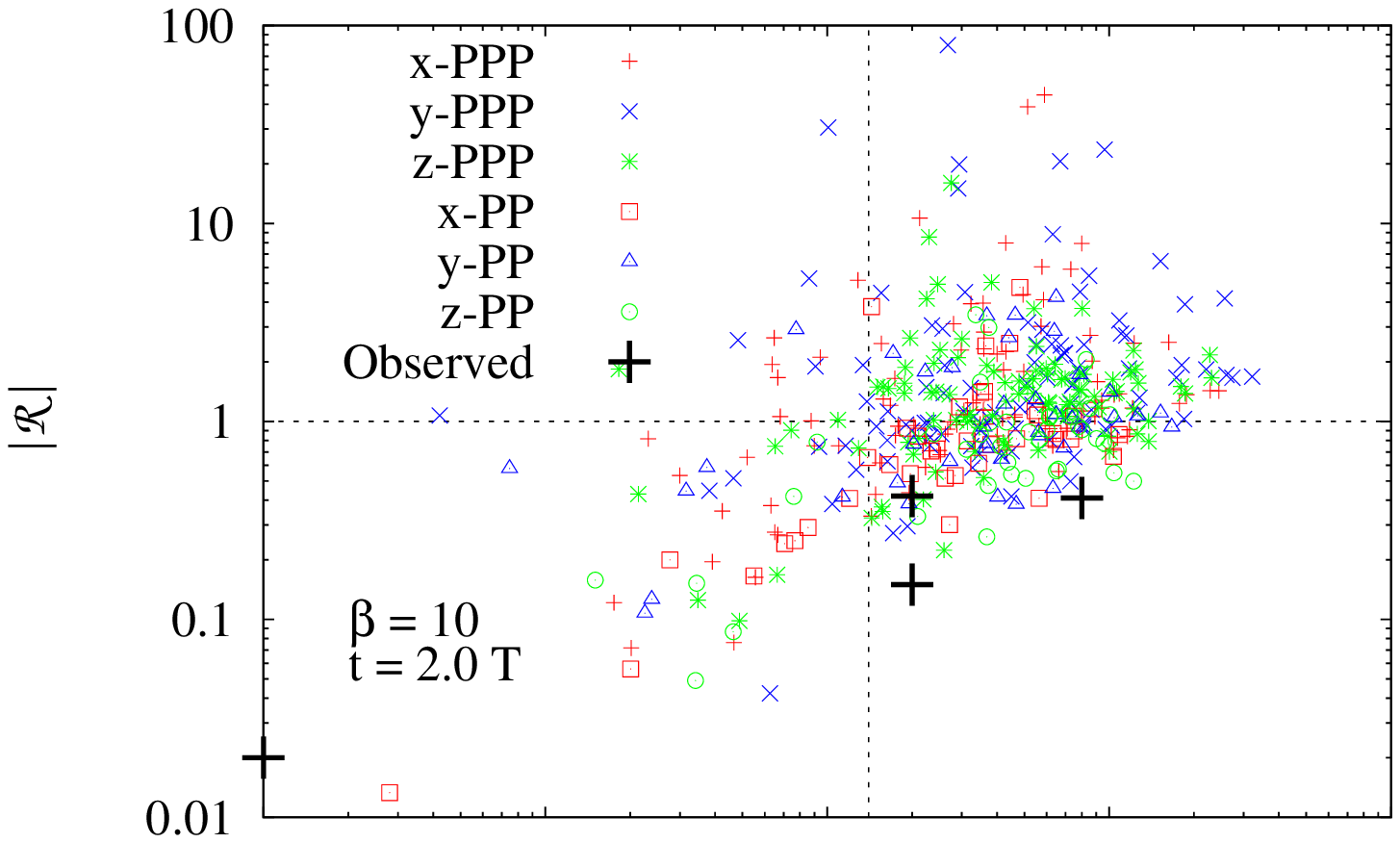}
\includegraphics[height=0.22\linewidth]{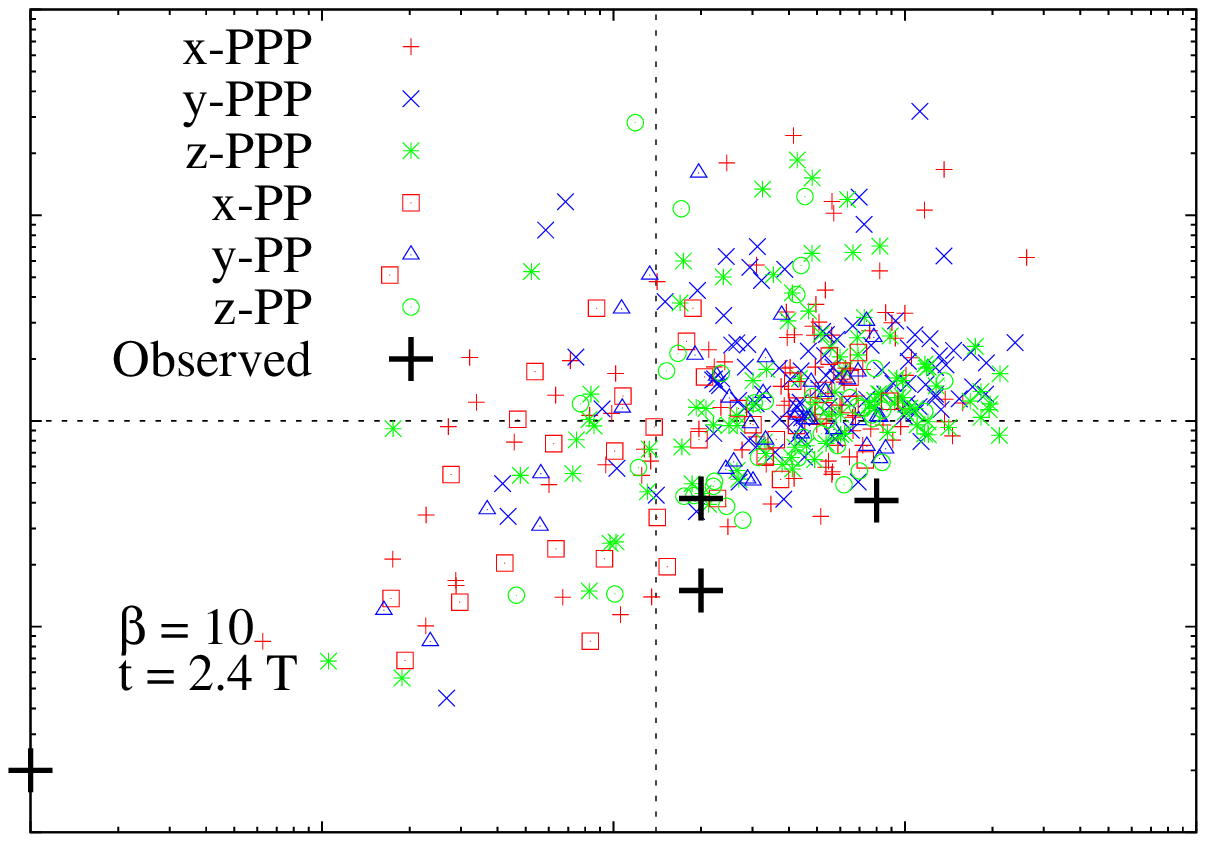}
\includegraphics[height=0.22\linewidth]{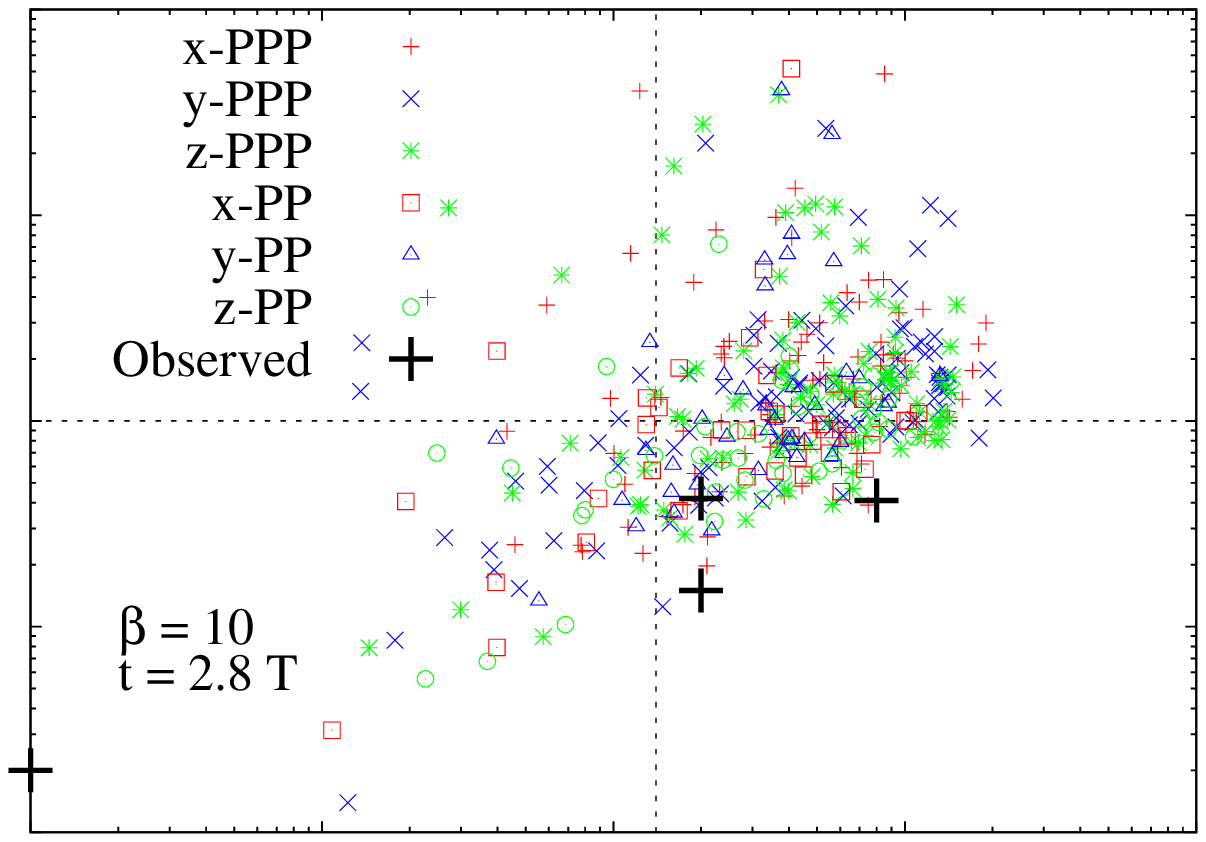}
}

\centerline{
\includegraphics[height=0.25\linewidth]{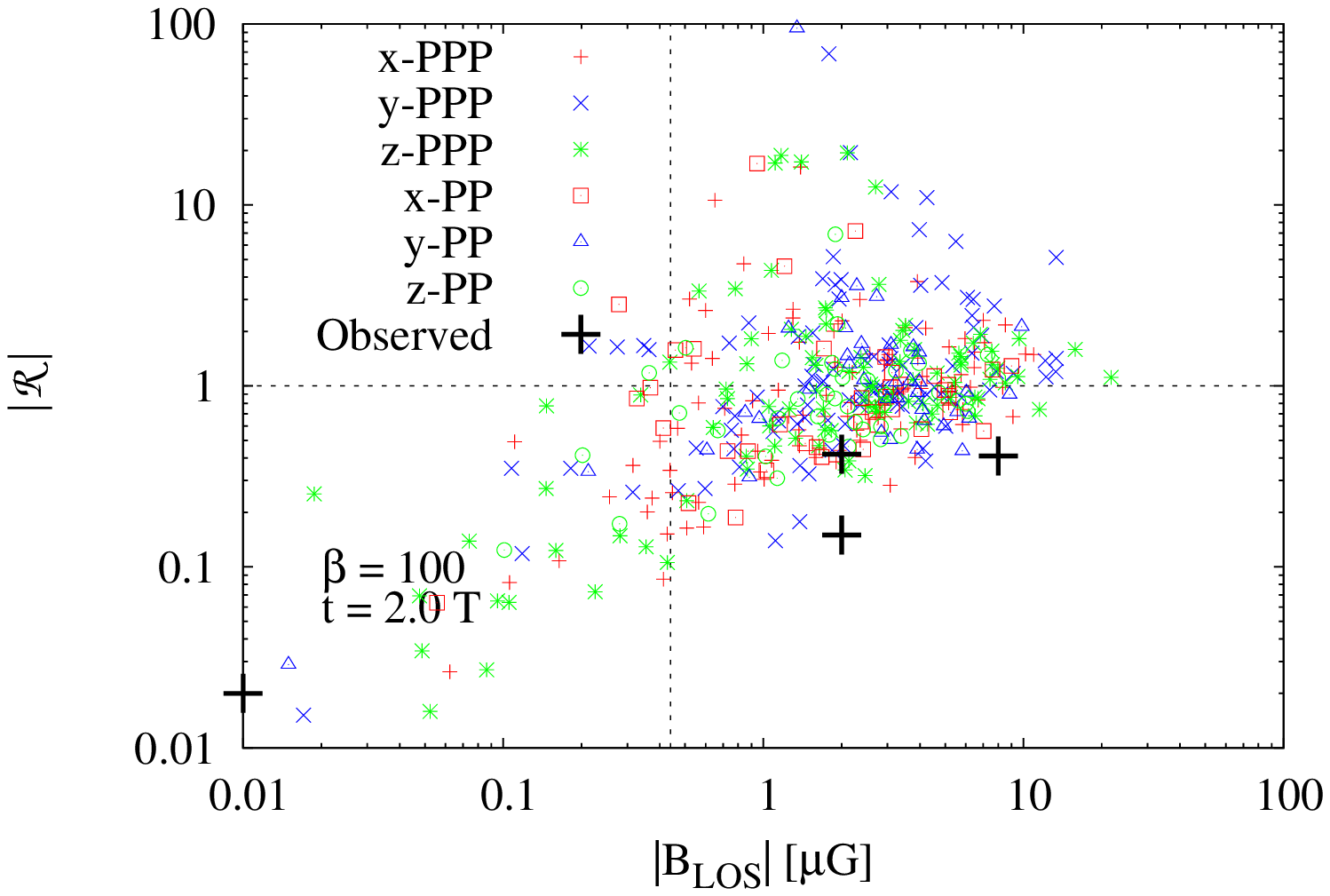}
\includegraphics[height=0.25\linewidth]{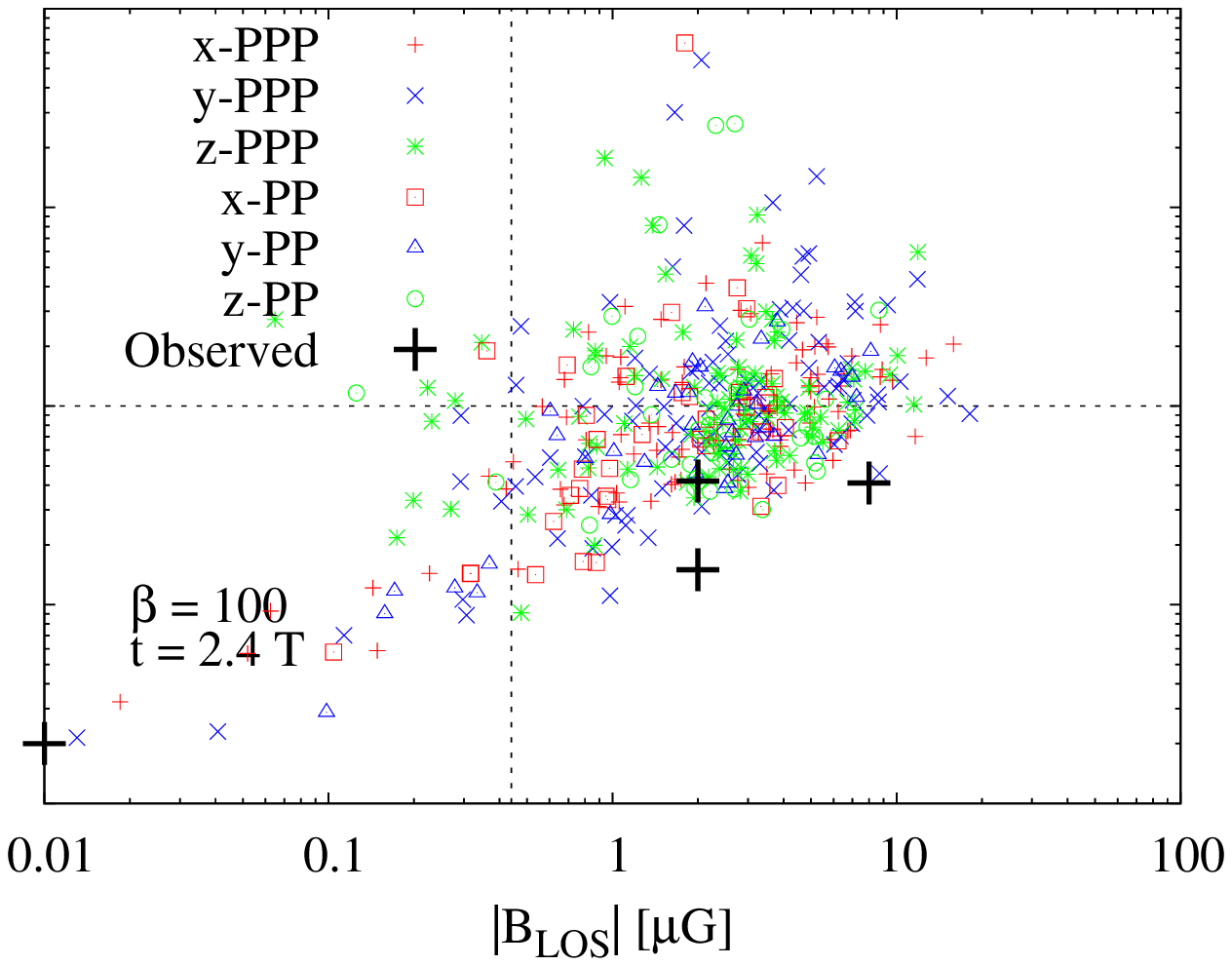}
\includegraphics[height=0.25\linewidth]{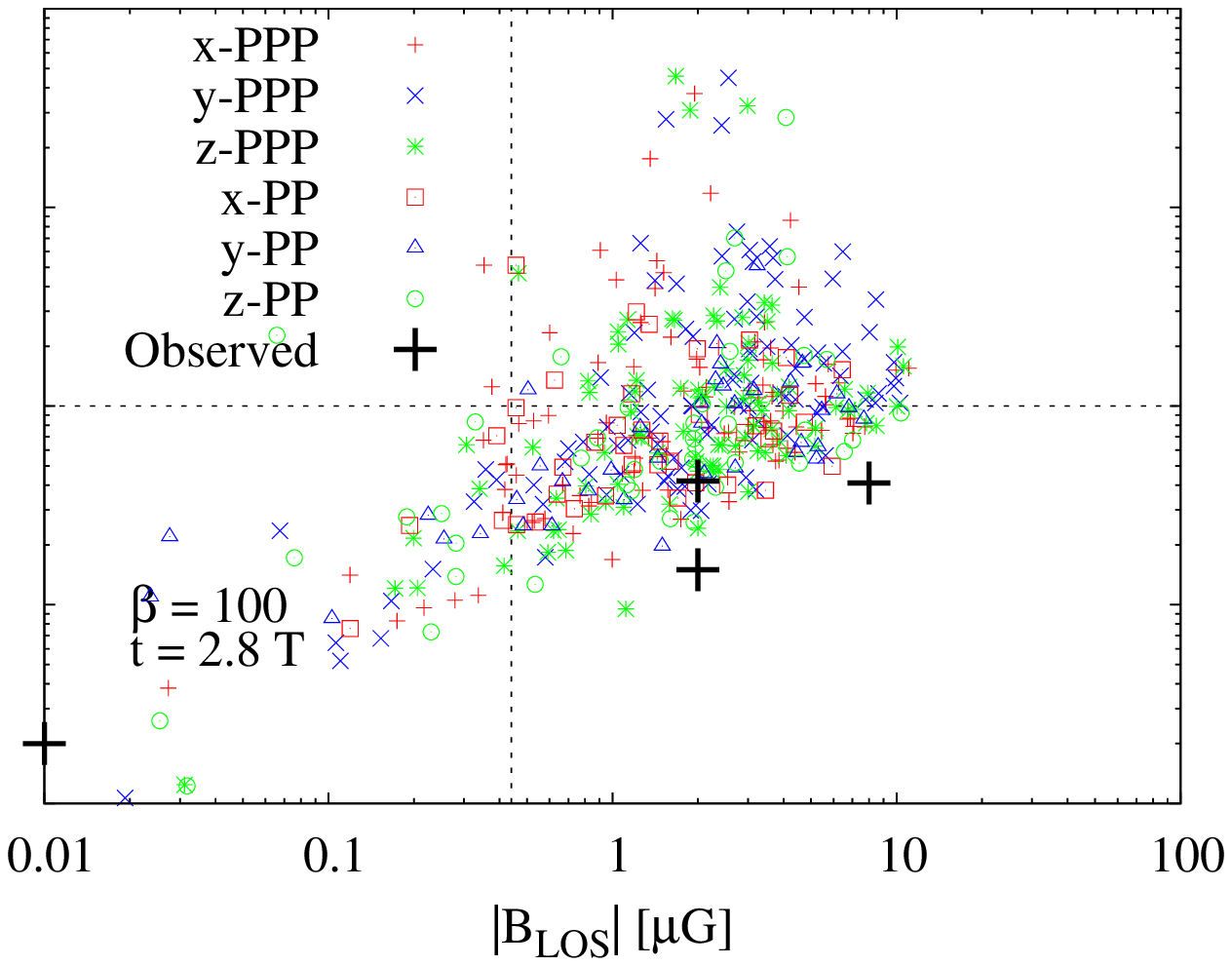}
}

\caption{Distribution of clumps in different LoS-directions i) for Position-Position-Position (PPP) and ii) for Position-Position (PP) measurements and observed cores by \citet{Crutcher}. From top to bottom: different values of $\beta_0$ ($\beta_0$ = 0.01, 0.1, 1, 10 and 100). From left to right: different time steps ($t = 2.0, 2.4$ and $2.8\,T$). The initial magnetic field strength for $\beta_0$ is marked with a vertical line. Plotted is the absolute value of $\R$ against the absolute value of the average of the magnetic field components for a given LoS. In general we observe a small value of $\Rb$ for small magnetic field strengths, that might be caused by field reversals. The stronger the magnetic field lines, the higher the value of $\Rb$. For PPP and PP configurations, as well as for the three different times, we get statistically the same distribution.}
\label{fig:scatter1}
\end{figure*}

\begin{figure*}
\centerline{
\includegraphics[height=0.224\linewidth]{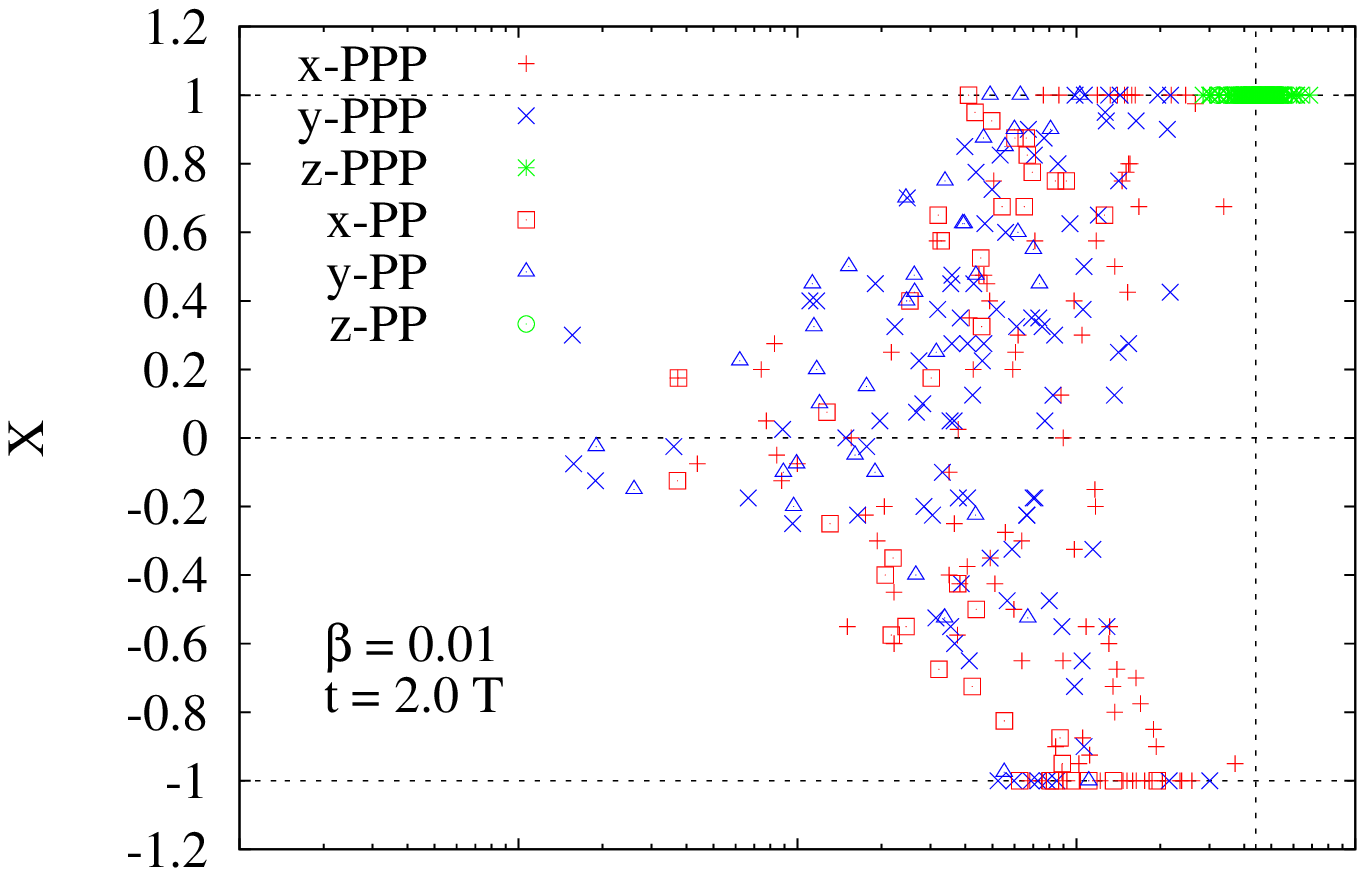}
\includegraphics[height=0.224\linewidth]{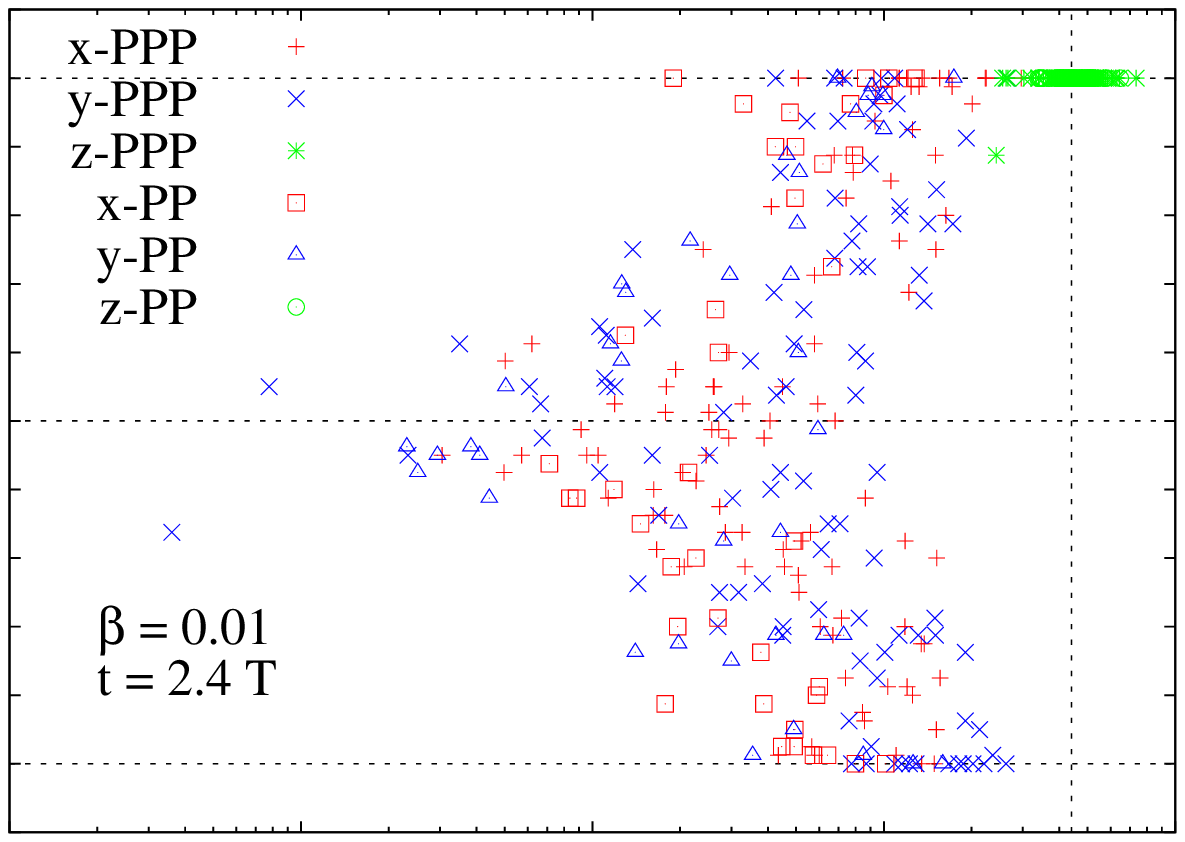}
\includegraphics[height=0.224\linewidth]{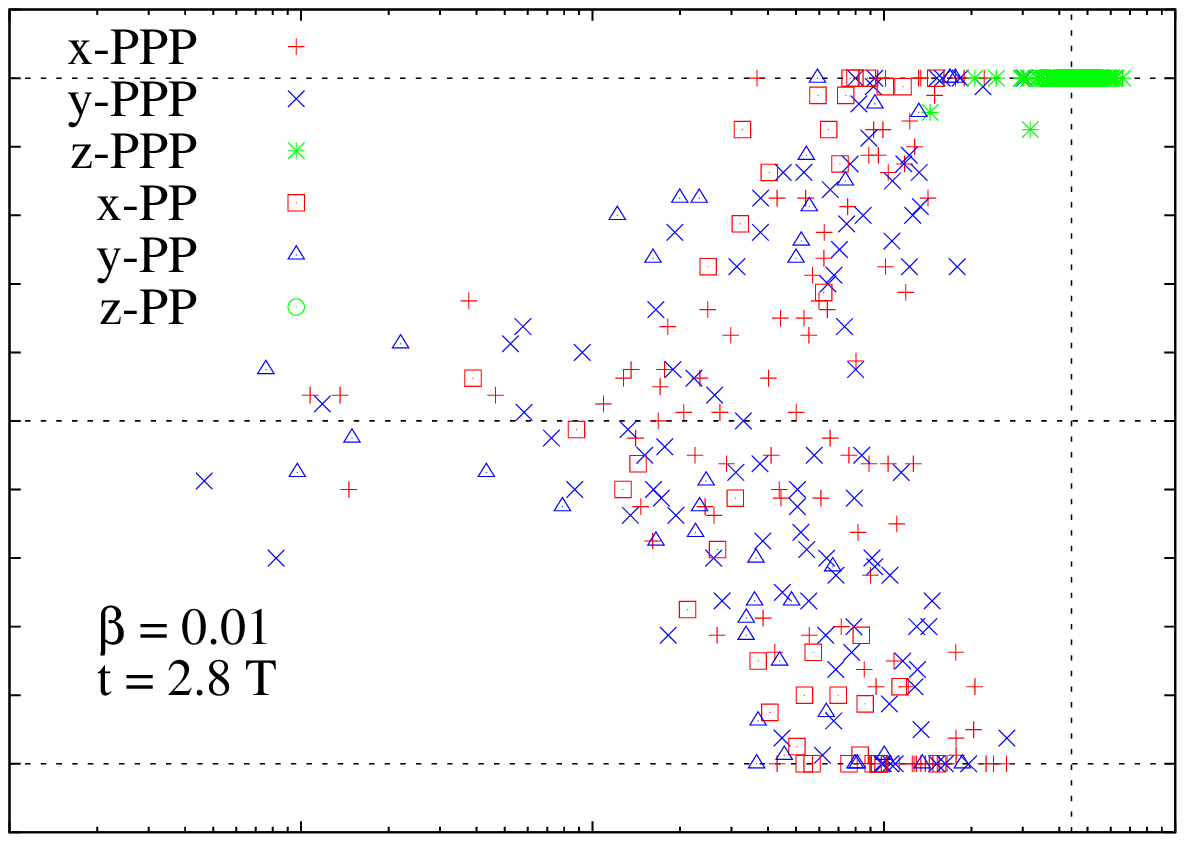}
}

\centerline{
\includegraphics[height=0.224\linewidth]{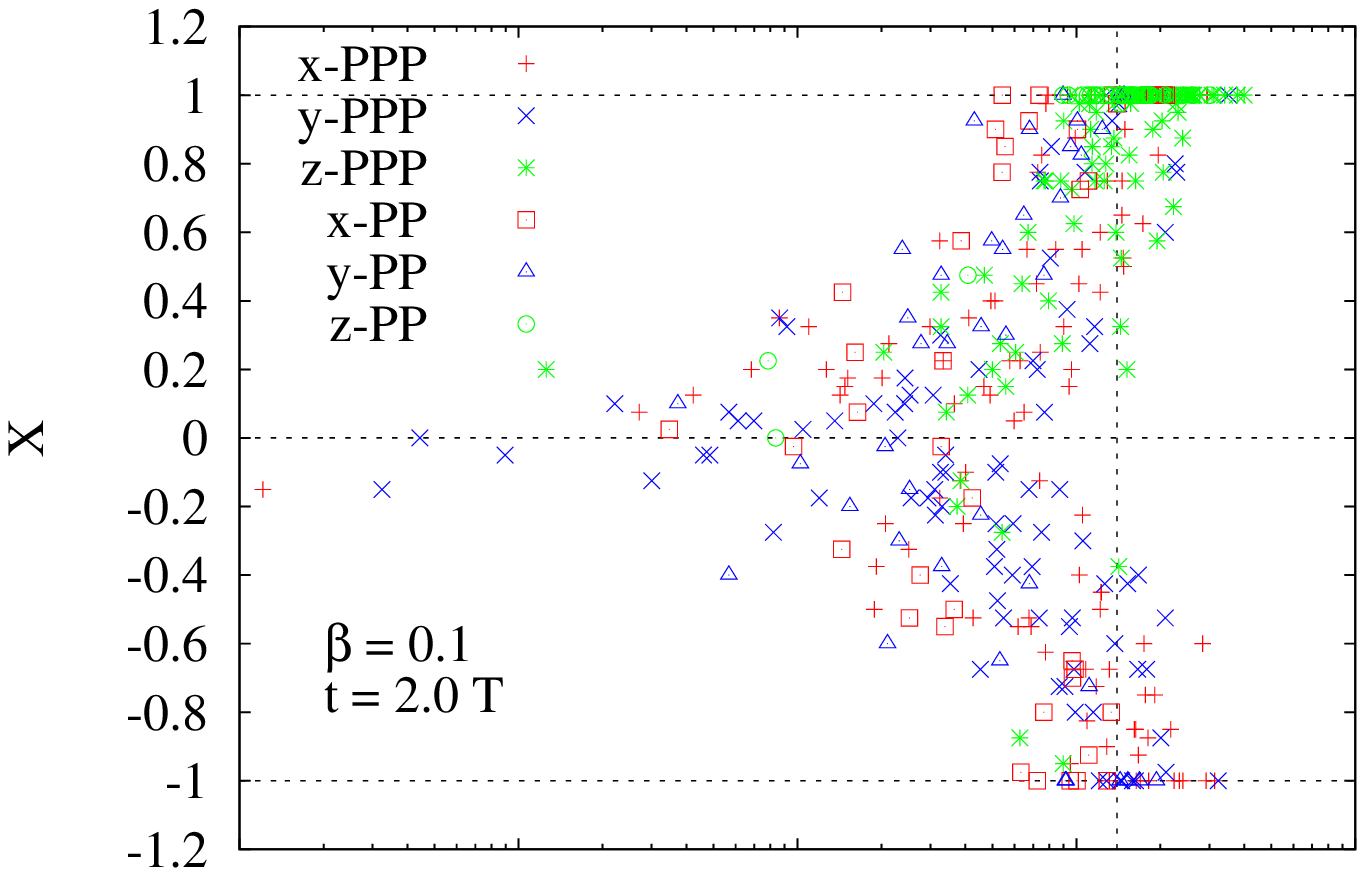}
\includegraphics[height=0.224\linewidth]{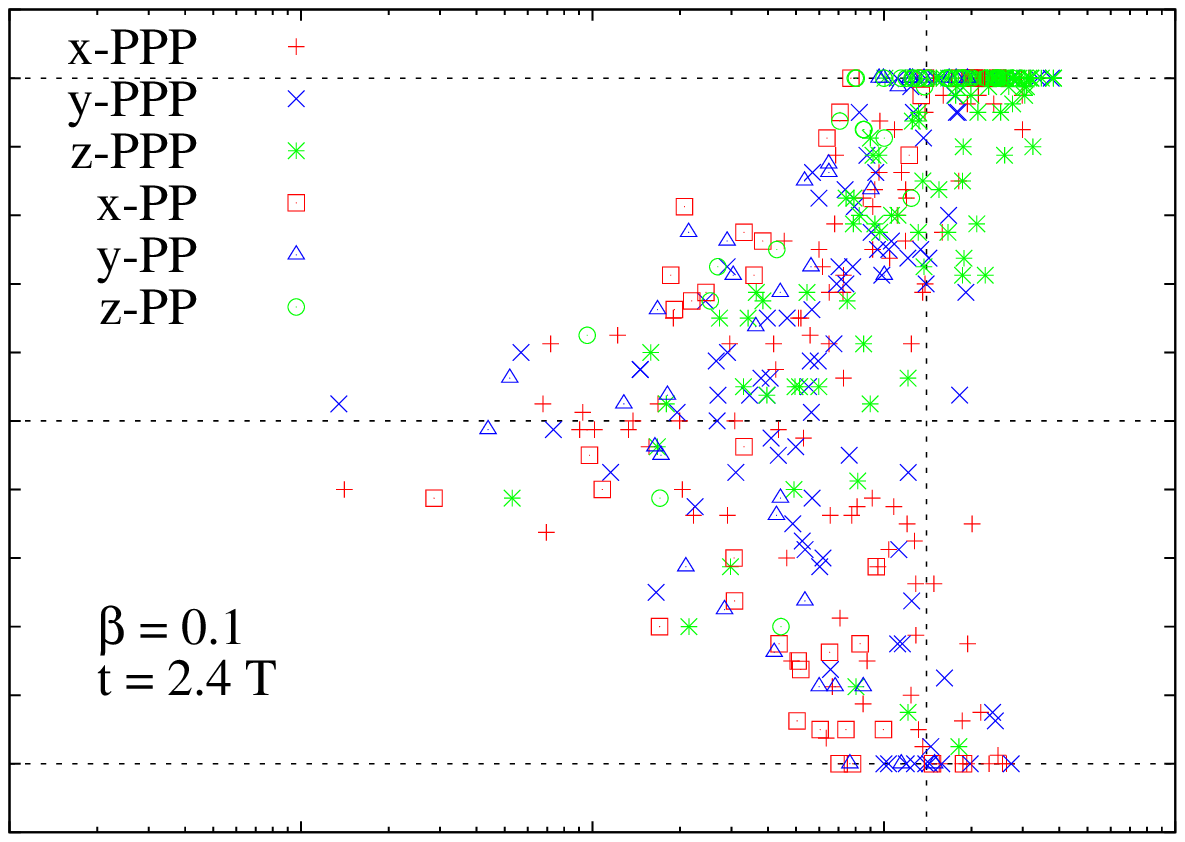}
\includegraphics[height=0.224\linewidth]{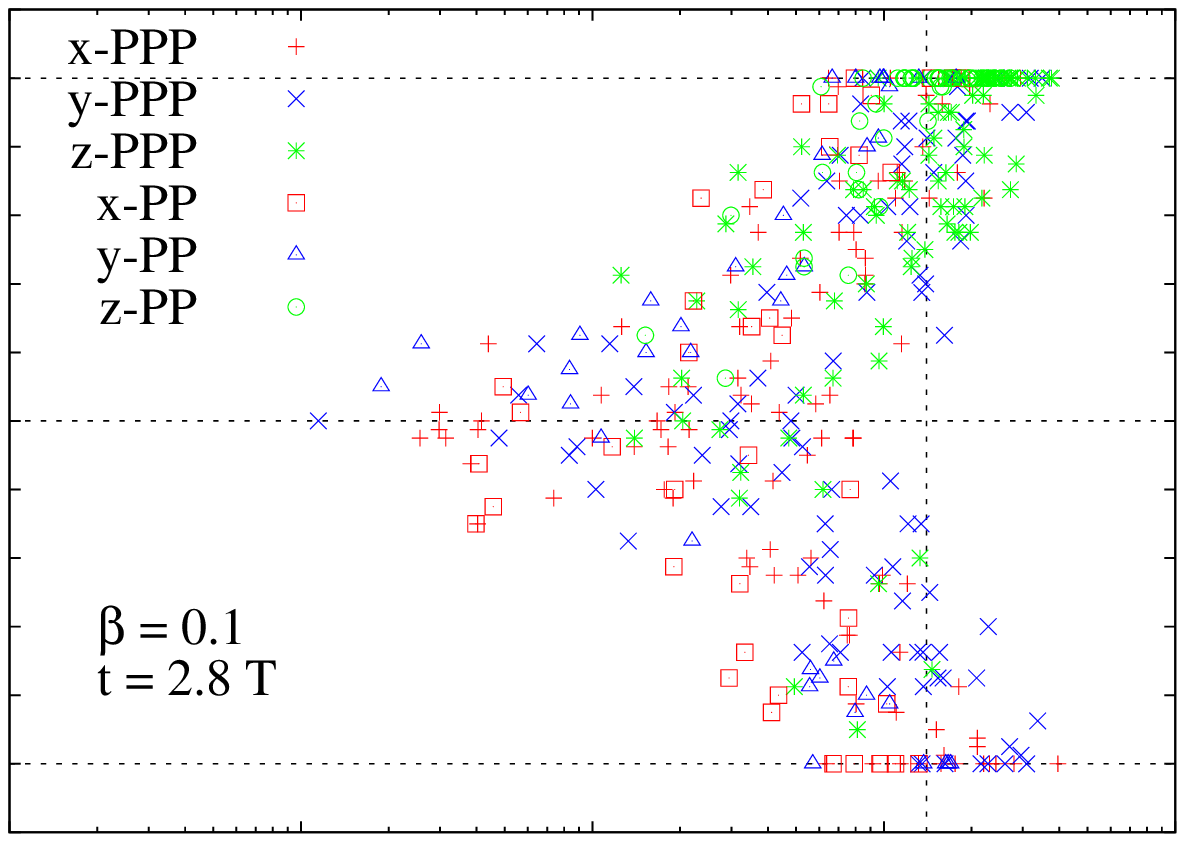}
}

\centerline{
\includegraphics[height=0.224\linewidth]{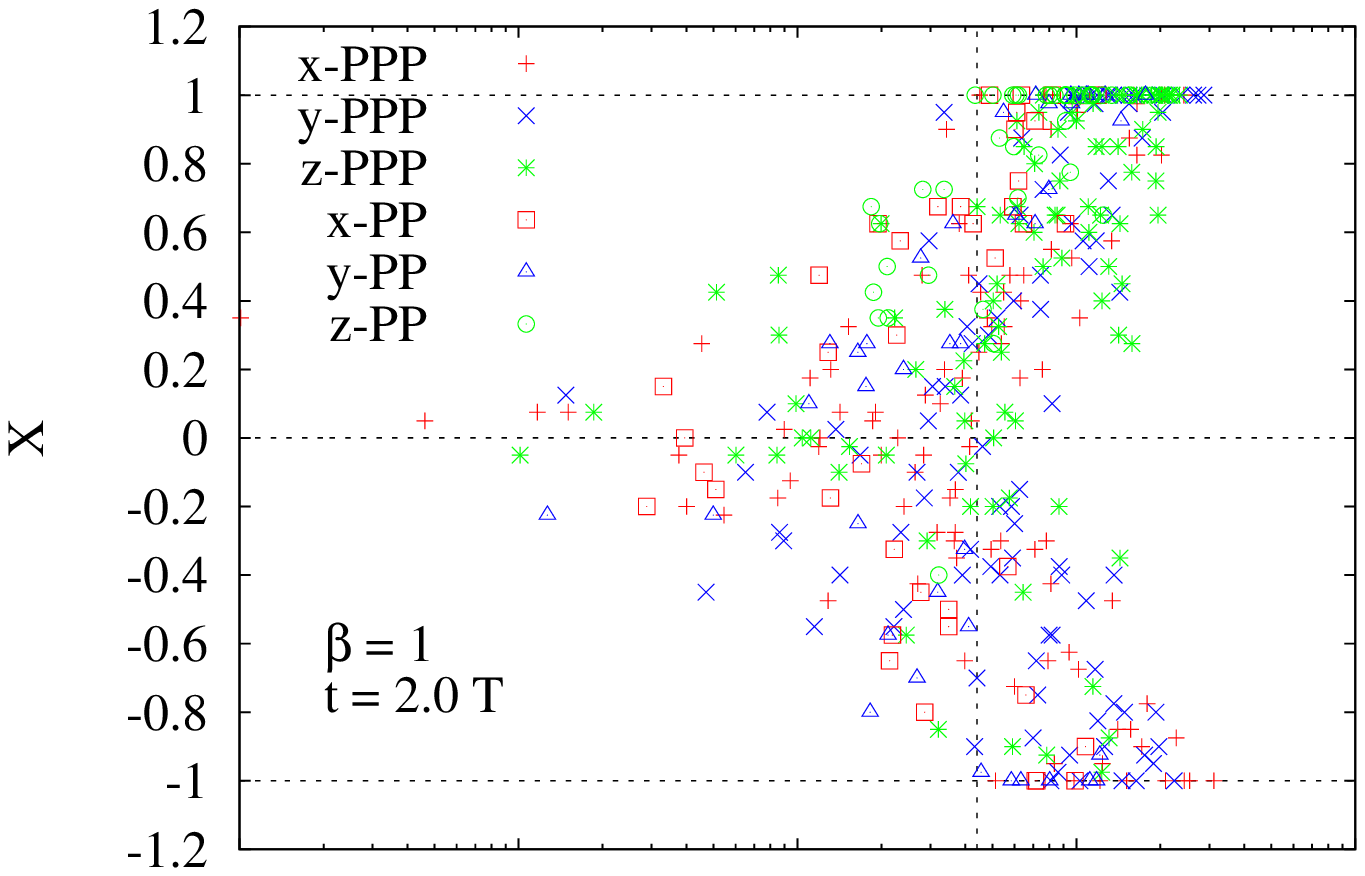}
\includegraphics[height=0.224\linewidth]{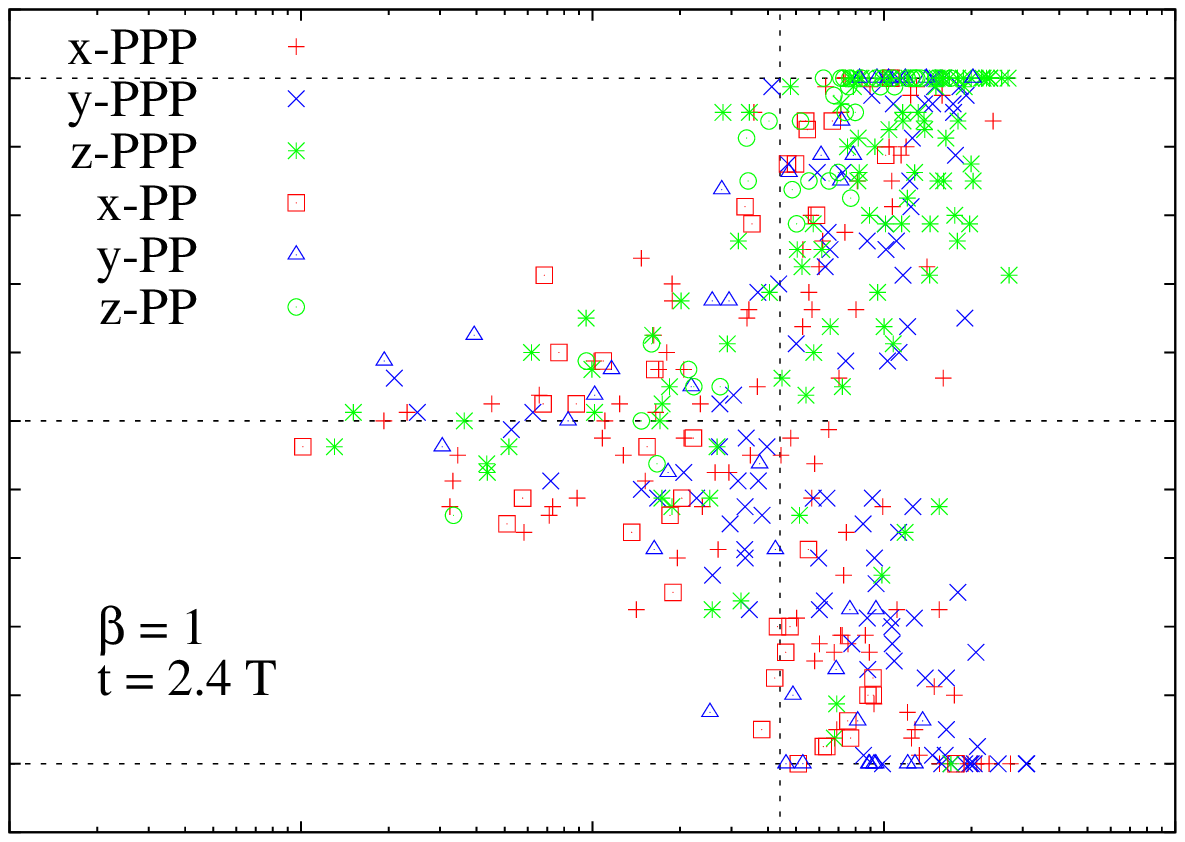}
\includegraphics[height=0.224\linewidth]{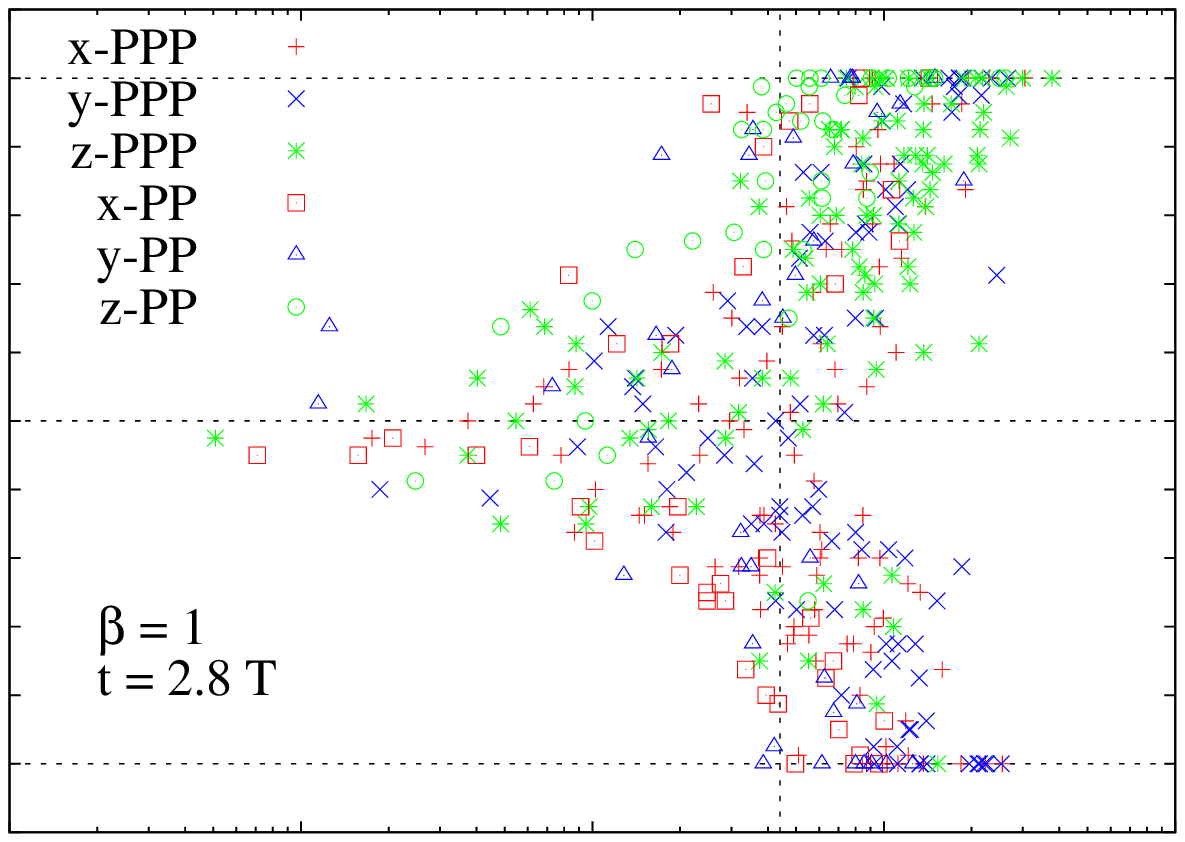}
}

\centerline{
\includegraphics[height=0.224\linewidth]{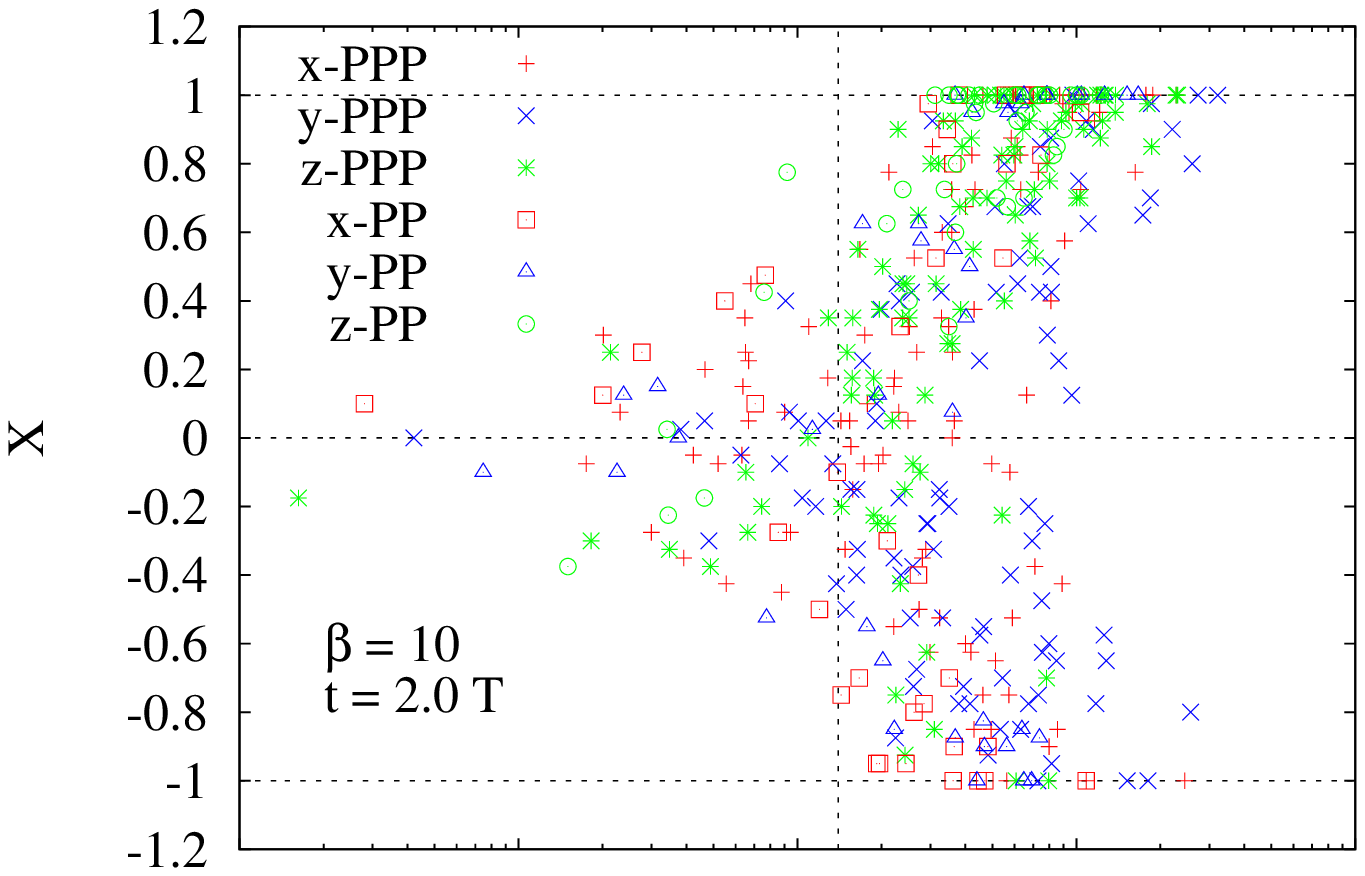}
\includegraphics[height=0.224\linewidth]{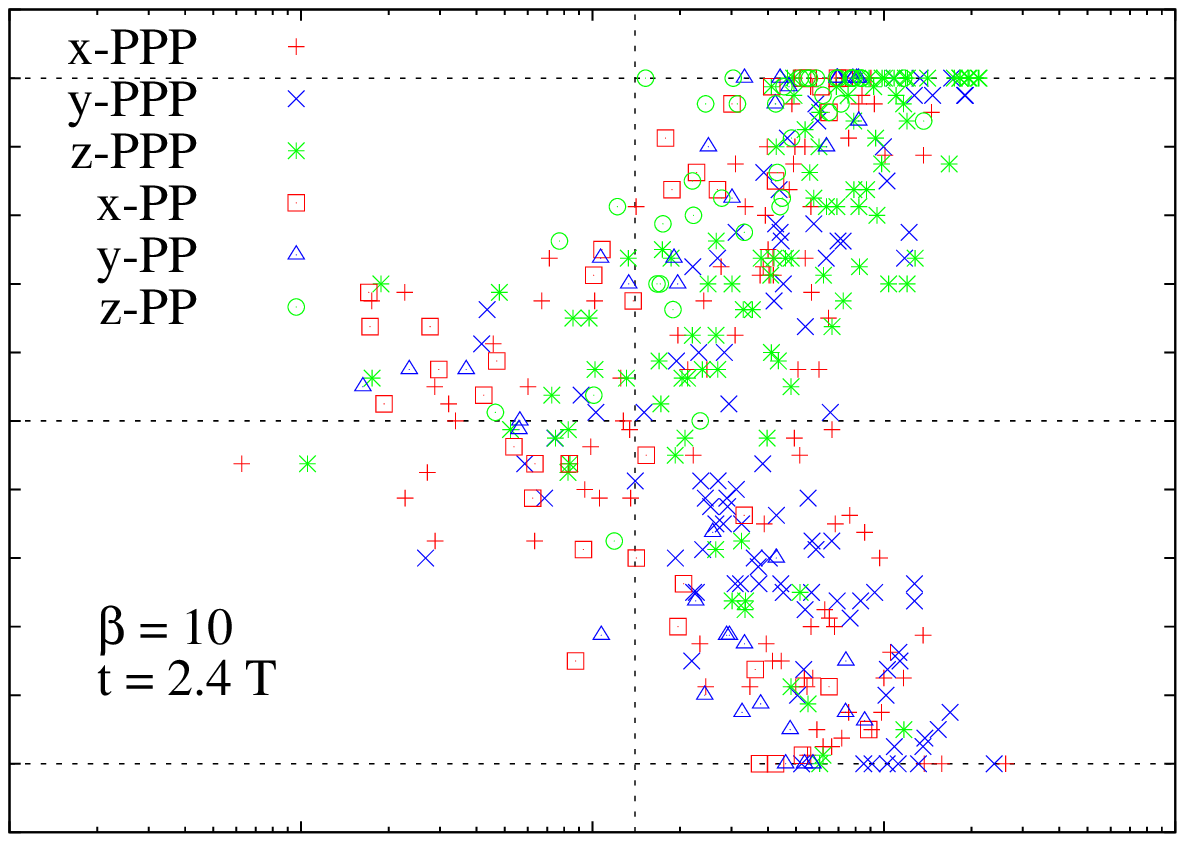}
\includegraphics[height=0.224\linewidth]{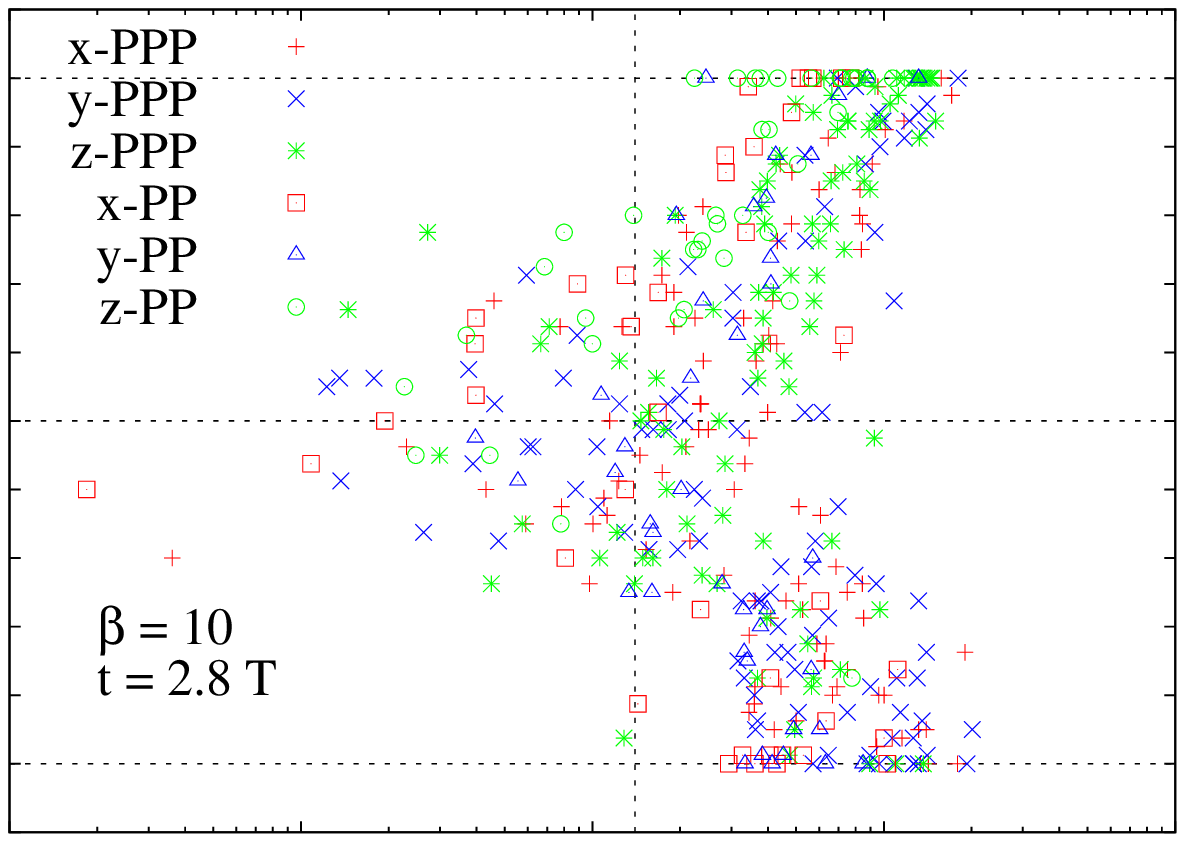}
}

\centerline{
\includegraphics[height=0.258\linewidth]{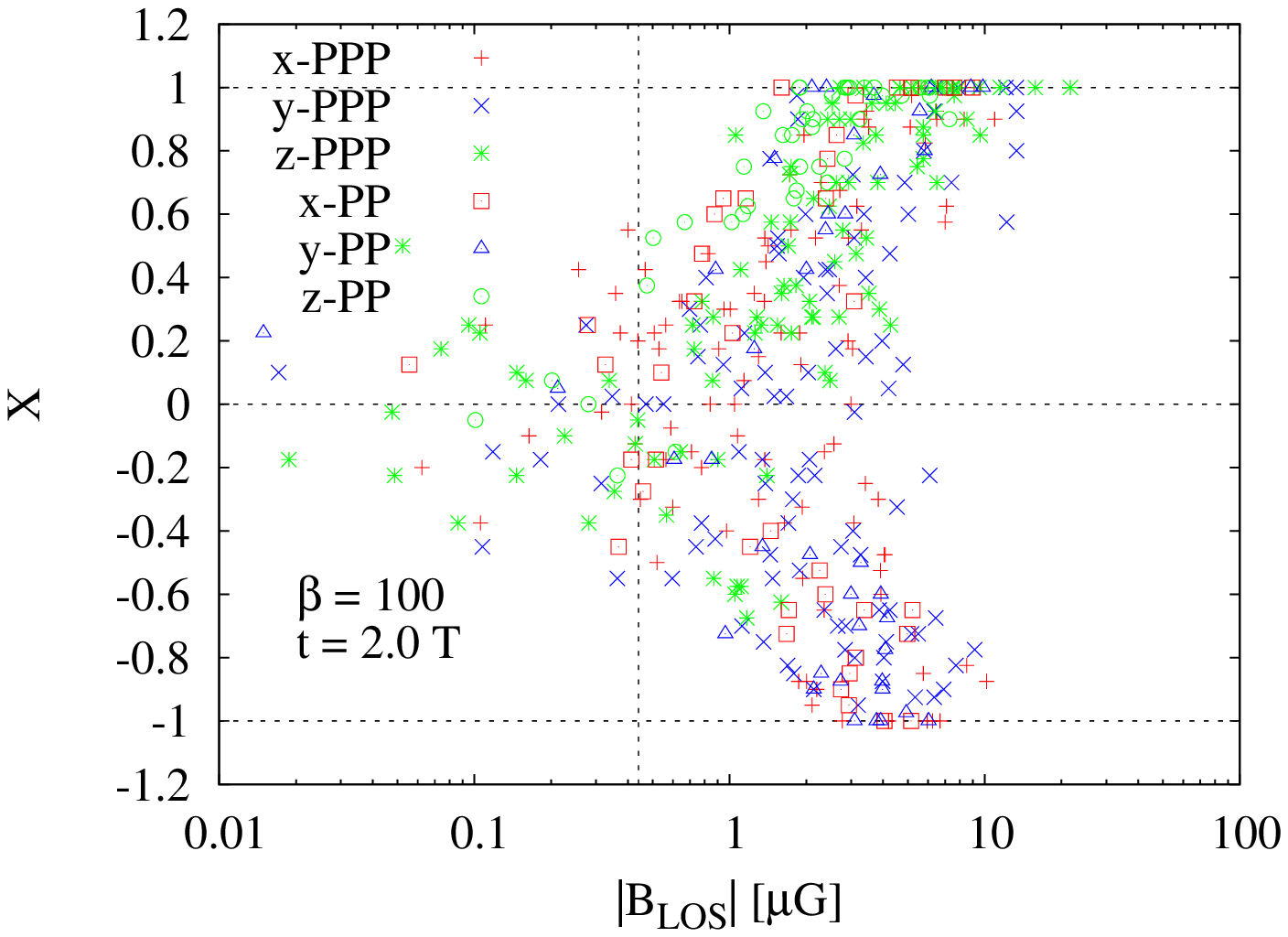}
\includegraphics[height=0.258\linewidth]{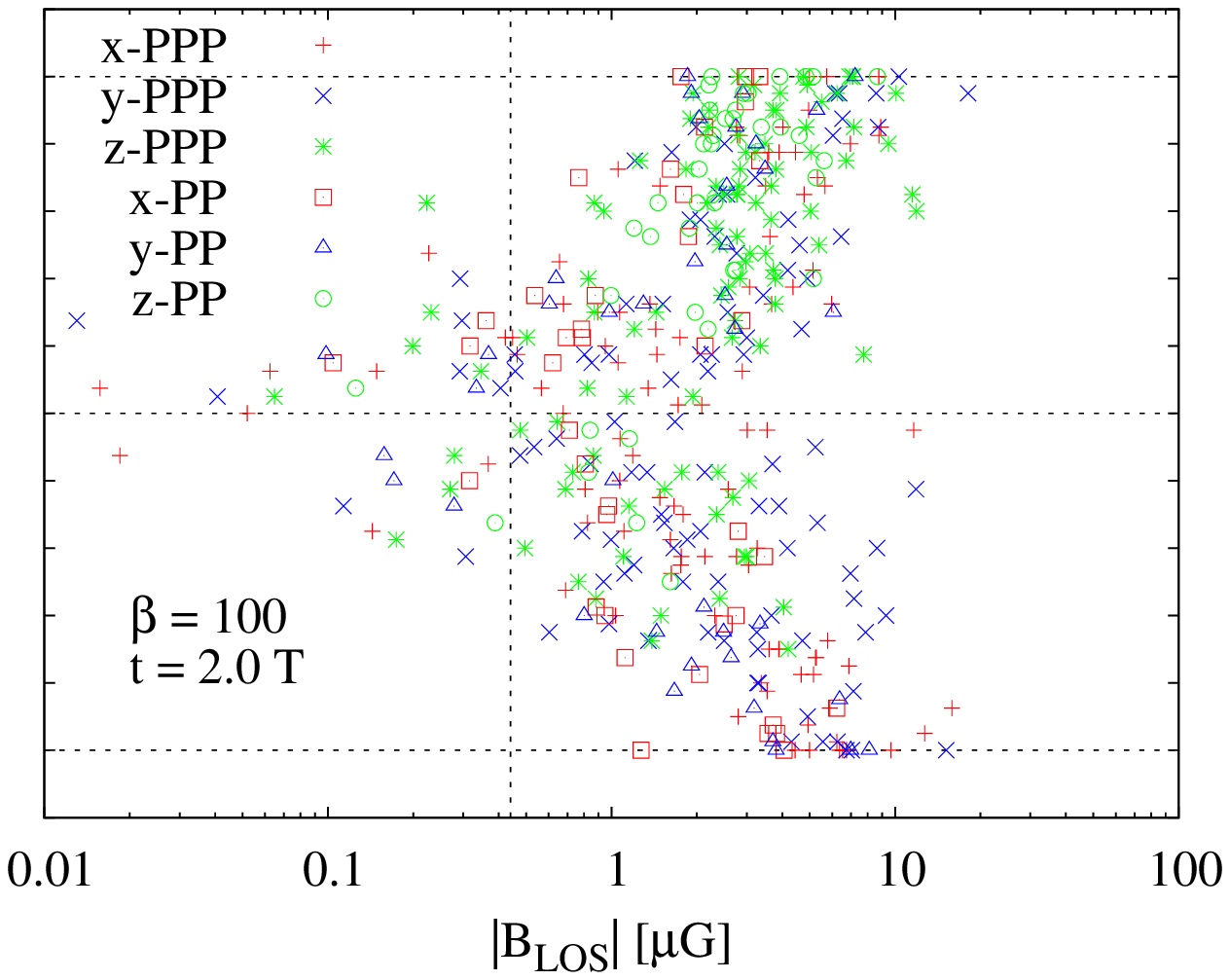}
\includegraphics[height=0.258\linewidth]{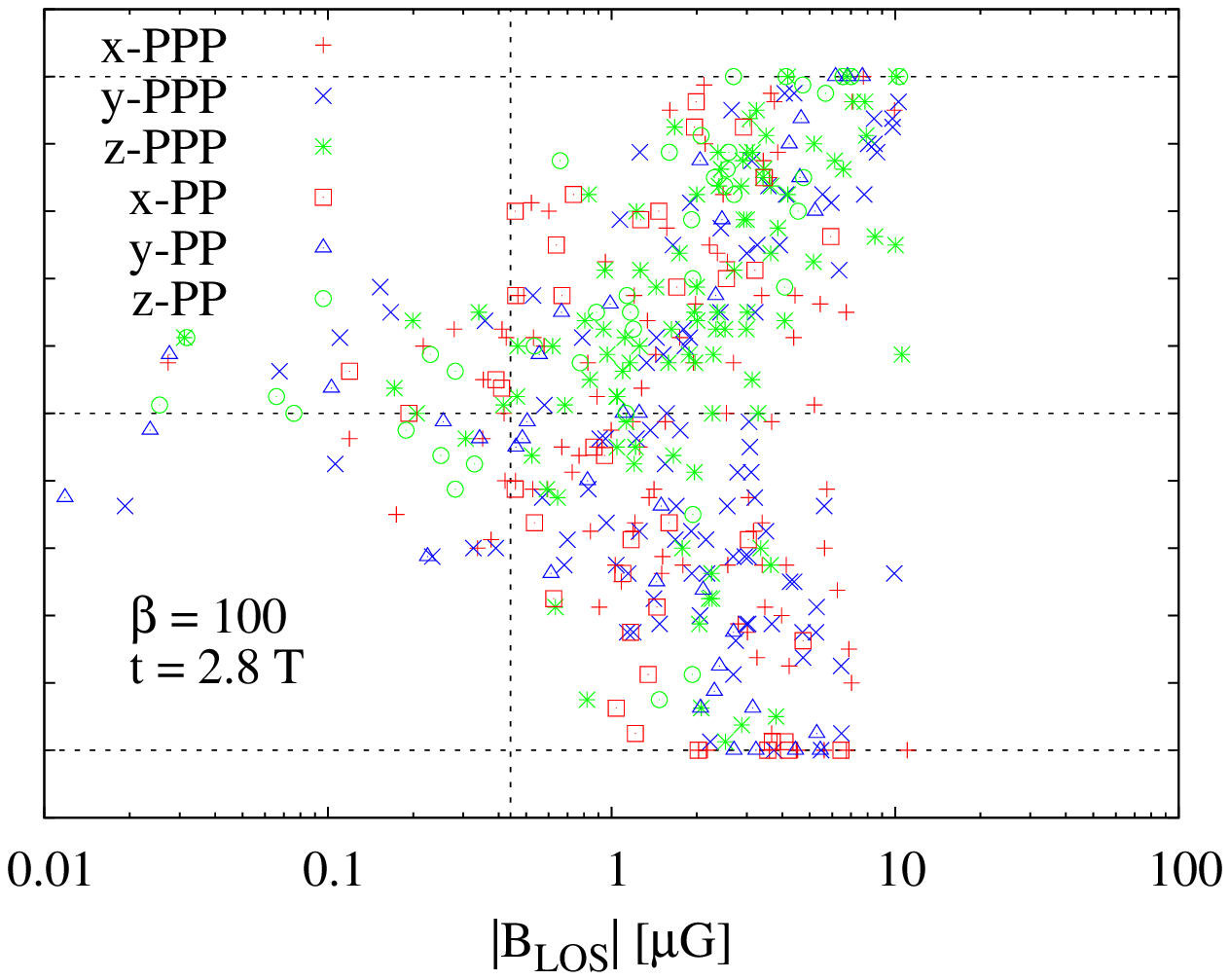}
}

\caption{Same as Figure ~\ref{fig:scatter1}, but for the number of field reversals, computed with eq.~(\ref{fieldreversals}).}
\label{fig:scatter2}
\end{figure*}

Let us now analyse the consequences of field reversals in our clumps on the behaviour of our statistical quantity $\R$ for our first analysis method. Figure~\ref{fig:scatter2} shows the corresponding amount of field reversals for each clump plotted in Figure~\ref{fig:scatter1}. Besides the fact that the distribution of clumps in Figure~\ref{fig:scatter2} qualitatively moves to lower magnetic field strengths as we go to higher values of plasma $\beta$, we notice that the standard deviation of $B$ and $\Rb$ observed in z-direction is getting bigger for lower magnetic field strength, i.e. for a higher plasma $\beta$. This is caused by the fact that a weaker field cannot resist as well against turbulence as strong magnetic fields, we therefore measure more field reversals as we go to higher plasma $\beta_0$ for the z-direction. We also observe that for small magnetic fields in the LoS (independent from any direction) we can identify more field reversals, that means clumps with values of $X \approx 0$.

\subsection{Comparison of two different methods of computing $\R$} \label{sec:twomethodsresults}

\begin{figure}
\centerline{
\includegraphics[width=1.0\linewidth]{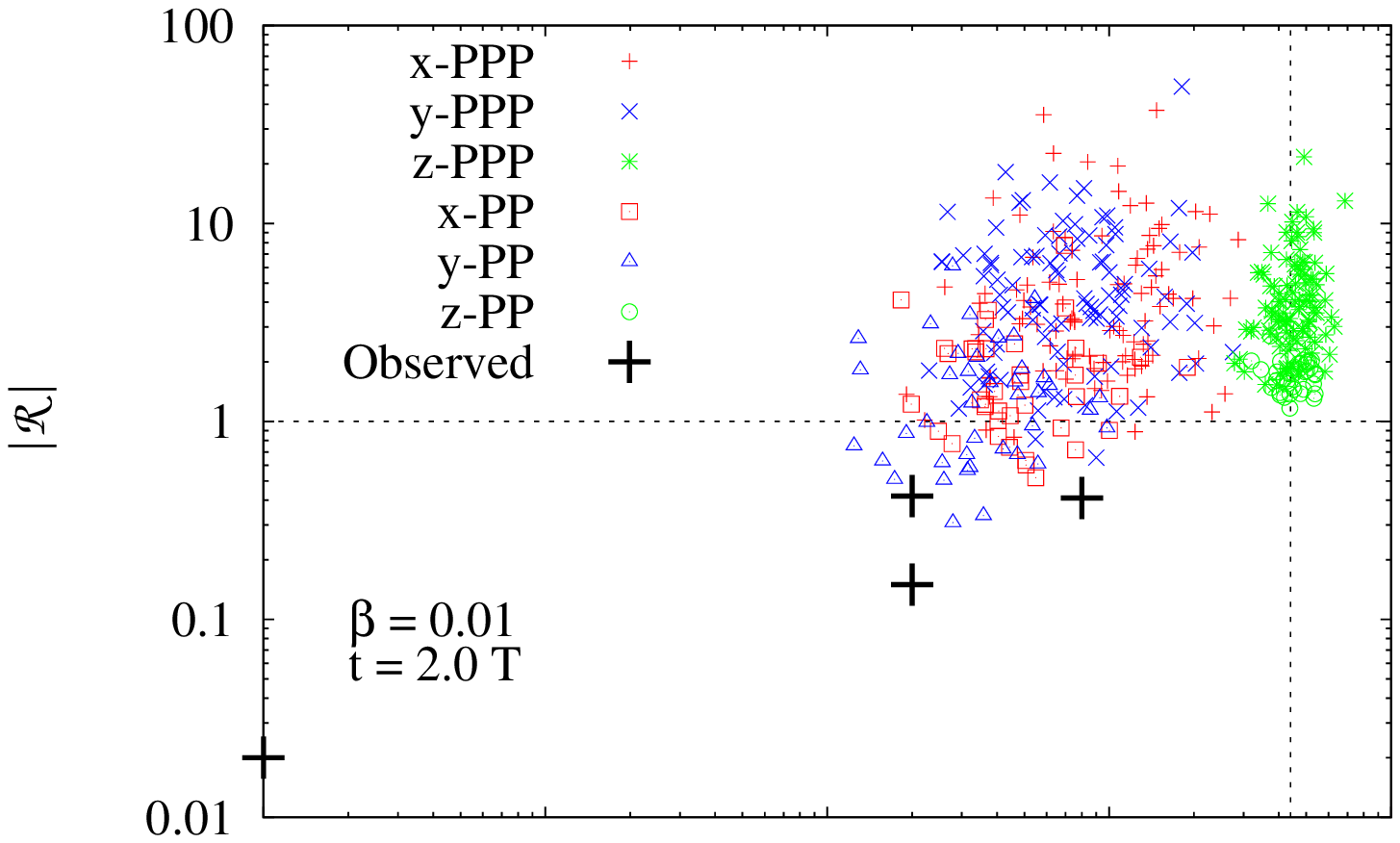} \\
}
\centerline{
\includegraphics[width=1.0\linewidth]{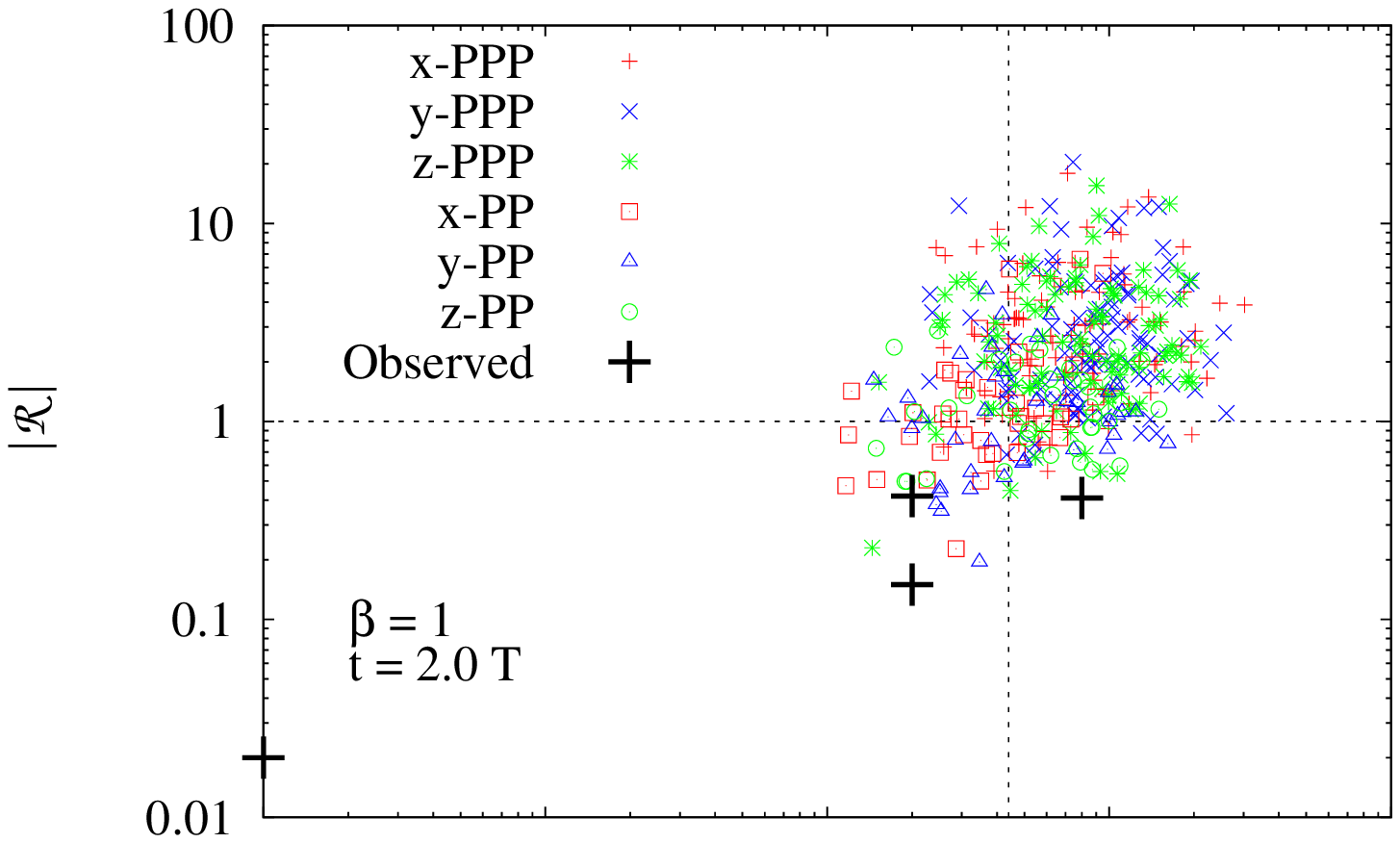} \\
}
\centerline{
\includegraphics[width=1.0\linewidth]{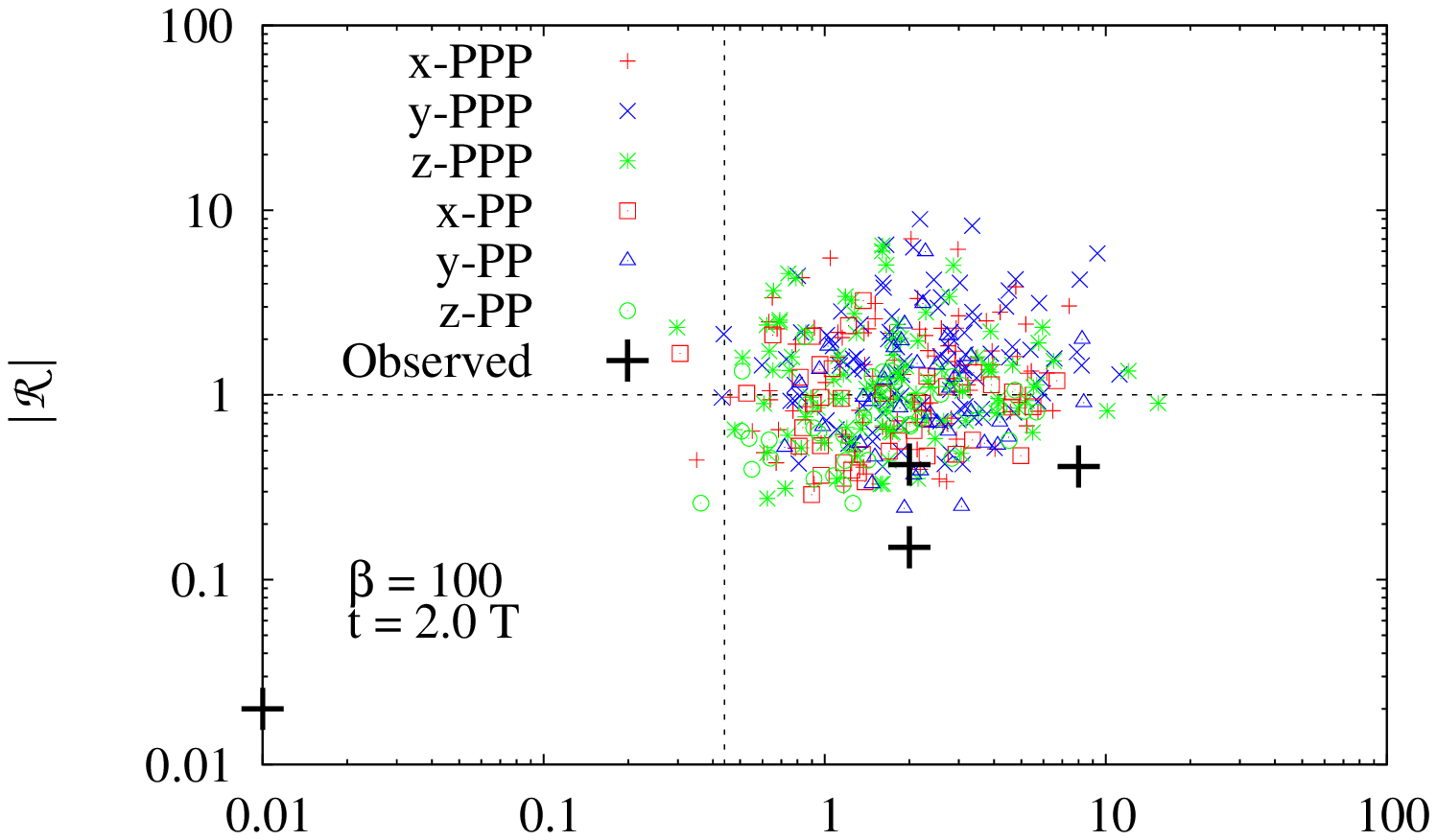} \\
}

\caption{Same as Figure~\ref{fig:scatter1}, but for the second analysis method describes in section~\ref{subsec:computeR}. Only $t = 2\,T$ and $\beta_0 = 0.01, 1$ and 100 are shown.}
\label{fig:scatter_log}
\end{figure}

In Section~\ref{subsec:computeR}, we have described two different methods of computing $\R$ for getting a statistical distribution of clumps in the $B-\R$-scatter plot. Depending on how we average our values we find some differences in the scatter plots. Figure~\ref{fig:scatter_log} shows the results when using our second analysis method for the same clumps used above. The observational measurements plotted in Figure~\ref{fig:scatter_log} are computed according to method 2 for the raw data given in \citet[][Tab.~1]{Crutcher}. Table~\ref{tab:method2} gives average values for all clumps in Figure~\ref{fig:scatter_log}. For both, PPP and PP, we again find mainly $\overline {\Rb} > 1$ for different $\beta_0$. As seen in Table~\ref{tab:values1} for our first analysis method, we also find a saturation of $\overline {\Rb} \approx 1$ for high $\beta_0$ for our second method. For example, the average PPP values for the x-direction at different $\beta_0$ are $\overline {\Rb} \approx 4.0, 3.4, 1.3$ (first method) and $\overline {\Rb} \approx 5.5, 3.6, 1.5$ (second method) for $\beta_0=0.01$, 1, and 100. The same trend can be observed for the other LoS-directions.

\begin{table}
\begin{tabular}{|l|l||l|l|l||l|l||l}
\hline\hline
LoS & $\overline {|B_{LoS}|}$ & $|\tilde{B}_{LoS}|$ & $\sigma_{|B|}$ & $\overline {\Rb}$ & $\tilde {\Rb}$ & $\sigma_{\Rb}$ \\
\hline
$\beta_0 = 0.01$ & & & & & & \\
\hline
PPP & & & & & &\\
\hline
 x & 10.6 & 10.0 & 5.8 & 5.5 & 3.3 & 6.1\\
 y & 8.2 & 6.9 & 4.7 & 5.7 & 4.3 & 5.7\\
 z & 45.5 & 45.7 & 7.8 & 4.6 & 3.8 & 3.0\\
\hline
PP & & & & & &\\
\hline
 x & 5.5 & 4.5 & 3.3 & 1.8 & 1.4 & 1.3\\
 y & 3.9 & 3.3 & 2.0 & 1.5 & 1.3 & 1.2\\
 z & 45.6 & 45.3 & 5.5 & 1.7 & 1.7 & 0.3\\
\hline
$\beta_0 = 1$ & & & & & & \\
\hline
PPP & & & & & &\\
\hline
 x & 8.6 & 7.3 & 5.3 & 3.6 & 2.8 & 3.0\\
 y & 9.6 & 8.9 & 4.9 & 3.6 & 2.5 & 3.2\\
 z & 9.0 & 8.6 & 4.9 & 3.2 & 2.4 & 2.6\\
\hline
PP & & & & & &\\
\hline
 x & 4.2 & 3.7 & 2.2 & 1.5 & 1.1 & 1.4\\
 y & 5.6 & 4.2 & 3.6 & 1.3 & 1.1 & 0.9\\
 z & 6.5 & 6.3 & 3.4 & 1.3 & 1.2 & 0.7\\
\hline
$\beta_0 = 100$ & & & & & & & \\
\hline
PPP & & & & & &\\
\hline
 x & 2.3 & 2.1 & 1.5 & 1.5 & 1.1 & 1.2\\
 y & 2.7 & 2.2 & 2.0 & 1.9 & 1.5 & 1.6\\
 z & 2.4 & 1.7 & 2.3 & 1.5 & 1.1 & 1.2\\
\hline
PP & & & & & &\\
\hline
 x & 2.0 & 1.4 & 1.5 & 1.0 & 0.9 & 0.6\\
 y & 2.5 & 2.2 & 1.6 & 1.2 & 0.9 & 1.0\\
 z & 1.8 & 1.5 & 1.3 & 0.9 & 0.8 & 0.4\\
\hline\hline
\end{tabular}
\caption{Mean, median and standard deviation for all directions for the magnetic field component and $\Rb$ for $256^3$ cells for $\beta_0 = 0.01, 1, 100$ for the PPP and PP case ($t = 2.0\,T$), computed with our second analysis method described in section~\ref{sec:methods}. All values of $B$ are given in $\mu$G.}
\label{tab:method2}
\end{table}

\section{Discussion}
\label{sec:discussion}

Here we propose two physical mechanisms to generate values of $\Rb\lesssim1$ and compare to existing studies. We distinguish between the strong and weak magnetic field limit. We note that with strong fields we do {\em not} refer to the classical model \citep{ShuAdamsLizano1987} where the energy density exceeds all other forms of energy and ambipolar diffusion is needed to form stars. Instead, by the strong magnetic field limit, we refer to field strengths that are close to saturation as expected by fundamental energy equipartition arguments \citep{Brandenburg}. Detailed simulations of the dynamo process under physical conditions, including magnetic field diffusion and viscous dissipation by \citet{FederrathEtAl2011PRL} predict the energy density of the field to be of order of 10 percent of the turbulent kinetic energy density for the typical transonic to supersonic Mach numbers in the interstellar medium.

\subsection{Strong magnetic field limit}
\label{subsec:stronglimit}

Molecular cloud cores are thought to form at the stagnation points of convergent flows \cite[see, e.g.,][]{hennebelle08, banerjeeetal09} in the turbulent interstellar medium, where the supersonic turbulent motions dominate the cloud on large scales \citep{Larson81} and become more coherent motions with subsonic velocity dispersion towards the cores centre \citep{ballesteros03, klessen05}. This transition to coherence typically occurs on scales of about $0.1\,$pc \citep{BensonAndMyers1989, goodman98, barranco98, AndreEtAt2007, LadaEtAl2008, BeutherAndHenning2009, SmithEtAl2009, FederrathEtAl2010, PinedaEtAl2010} which is comparable to the diameter of our simulated clumps of $0.08\,$pc (see Table \ref{tab:core_prop}). Because of the more coherent flow pattern, the magnetic field lines are less strongly twisted in the central core compared to the outer envelope. In addition, the field strength in the core is larger than in the envelope due to the compression of the field lines as the density increases. Both aspects make field reversals more likely to occur in the envelope, where magnetic field lines can cancel out, than in the inner core. As a result we measure $\Rb\lesssim1$. In principle $\Rb\lesssim1$ could also result from geometrical effects and observational bias \citep{VazquezEtAl2005,Crutcher}.

\subsection{Weak magnetic field limit}
\label{subsec:weaklimit}

Our statistical analysis of turbulent core formation in section~\ref{sec:results} yielded average values of $\Rb\gtrsim1$. Only for the smallest considered initial field strengths, we found a significant number of cores with $\Rb\lesssim1$, but still with an average $\Rb$ very close to unity. However, as discussed above, these calculations focus on a regime where the magnetic field is close to the saturation level. It is also important to consider systems where the field strength is orders of magnitudes below this value. To do so, we obtained simulation data from \citet{SurEtAl2010} and \citet{FederrathEtAl2011} where the initial field strength was only $B = 10^{-9}\,$G and then was amplified by a factor of $10^4$ by the small-scale turbulent dynamo driven by gravitational collapse \cite[see ][, for a general discussion of the energetics of turbulence generated by gravitational contraction]{klessen10}.

\begin{figure}
\centerline{\includegraphics[width=1.0\linewidth]{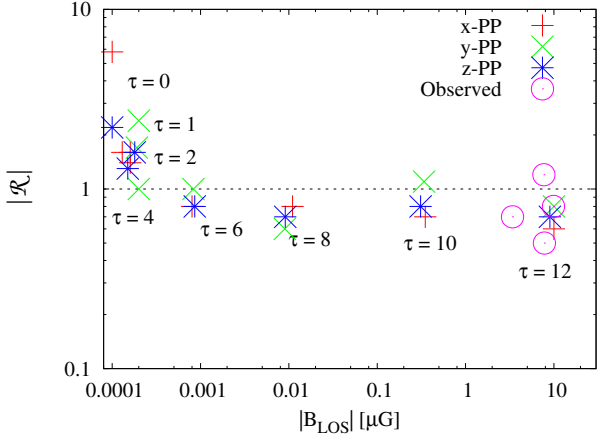}}
\caption{Time evolution of $\Rb$ for the turbulent, weakly magnetised, collapsing Bonnor-Ebert sphere of \citet{SurEtAl2010}. The initial turbulent phase, $\tau\lesssim4$, leads to a decrease of $\Rb$ to around unity, due to field reversals. The magnetic field gets further amplified by the \emph{gravity-driven, small-scale turbulent dynamo} \citep{FederrathEtAl2011} during the subsequent collapse phase, $\tau\gtrsim4$, which leads to $\Rb<1$.}
\label{fig:BESphere}
\end{figure}

Imagine a very weak initial magnetic field that is amplified due to the stretching, twisting, and folding of magnetic field lines, the so-called \emph{small-scale dynamo} \citep[see][for a review]{Brandenburg}. The turbulent dynamo is a process by which turbulent kinetic energy is converted into magnetic energy, by packing field lines closer together. This process always works, if turbulent kinetic energy is injected into the system, and as long as the magnetic energy is still smaller than the kinetic energy, from which the dynamo feeds. \citet{SurEtAl2010} and \citet{FederrathEtAl2011} showed recently that turbulent dynamo amplification also works in a collapsing, magnetised core. Thus, even if $\Rb>1$ initially, we expect that $\Rb$ becomes smaller than one, as the magnetic field gets amplified in the core. Figure~\ref{fig:BESphere} shows the time evolution of $\Rb$ for the turbulent, magnetised core studied in \citet{SurEtAl2010} and \citet{FederrathEtAl2011}. The core starts off with a very weak initial magnetic field and $\Rb=5.8$ at time $\tau=0$. The initial turbulence prevents the core from collapsing immediately, and makes $\Rb$ decrease to around unity at $\tau\approx4$, when core collapse sets in. This initial decrease of $\Rb$ is due to field reversals as explained for the turbulent models in section~\ref{sec:fieldreversals}. During the collapse phase, however, the magnetic field is amplified further, and for $\tau\gtrsim4$, we find values $\Rb<1$, as expected, due to the dynamo-amplified flux in the centre of the core. We conclude that magnetic field amplification during gravitational collapse, as reported in \citet{SurEtAl2010} and \citet{FederrathEtAl2011}, can be an important mechanism to explain the low values of $\Rb<1$ found in the core observations by \citet{Crutcher}.

\subsection{Relation to existing studies in the literature}

On average, our results seem to show the opposite trend found by \citet{Lunttila}, who observed that most clumps have values of $\Rb < 1$ for larger magnetic fields and values of $\Rb > 1$ for lower magnetic fields. However, we find approximately 70--90$\%$ of our clumps for $\beta_0 = 0.01$ and 40--60$\%$ of our clumps for $\beta_0 = 100$ having values of $\Rb > 1$, despite of a very broad 1$\sigma$-interval. Thus, for small magnetic field strengths, we find more clumps with small $\Rb$ than for higher field strengths. We interpret this as being a result of field reversals, which naturally leads to smaller values of $\Rb$, if the field is more easily tangled. In addition to that, small fields can be amplified by the small-scale turbulent dynamo action, again leading to a predominately smaller values of $\Rb$. Both aspects lead to the trend we observe in our statistical analysis, i.e., that $\Rb$ is smaller for smaller magnetic fields.

This trend is basically in agreement with the \citet{Crutcher} cores although only four clumps are analysed. Any trend seen in their observations might thus be due to low-number statistics and does not necessarily reflect average clump properties. Similar holds for the average values. The average values of $\Rb$ in our simulations are mostly larger than 1, but \citet{Crutcher} only observes values $\Rb\lesssim1$, with some dependency on the analysis method (see sec.~\ref{sec:twomethodsresults}). The question remains whether this is caused by a physical or statistical effect, because of lack of observational data. In fact, our results show that the quantity $\Rb$ is not necessarily a good statistical measure to distinguish between ambipolar diffusion theory and turbulent theory of star formation, because of the large distribution of $\Rb$ with values of $\Rb$ both larger and smaller than unity. In fact, we find that most of our clumps have values with $\Rb\gtrsim1$, even in a turbulent environment \emph{without} ambipolar diffusion acting in our simulations.

One might argue that numerical diffusion could have had a similar effect as ambipolar diffusion, and that the fact that we obtain mean values $\overline{\Rb}\gtrsim1$ is a numerical effect. This, however, can be safely excluded as shown in Appendix~\ref{app:resol}. There, we show that the mean properties of our clumps do not show a systematic dependence on resolution. In addition, \citet[][App.~C]{FederrathEtAl2011} explicitly measured the numerical diffusion of the present MHD scheme. We thus find that numerical effects have no impact on our general conclusions.

A difficulty of interpreting these results is the projection of clumps along the LoS-axis for the PP case. The projection leads to blending of structures in the LoS, so that a clump found in the column density map (PP) could also be a superposition of more than one object (or filament) that are spatially separated from one another \citep{Ballesteros-Paredes}. This is the reason why we also analysed the PPP case and compared both.

Finally, we would like to mention the observational challenges of measuring $B_{LoS}$ in the envelope of a clump and the recent discussions in this context \citep[see, e.g.,][]{MouschoviasAndTassis2009,CrutcherEtAl2010,MouschoviasAndTassis2010}.

\section{Summary and Conclusions} \label{sec:summary}

We have analysed MHD simulations of driven, supersonic turbulence with different initial magnetic field strengths ($\beta_0=0.01$, 0.1, 1, 10, and 100) at different turbulent turnover times, $t=2.0$, 2.4, $2.8\,T$. Our aim was to analyse the statistics of the parameter $\R = (M/\Phi)_{core} / (M/\Phi)_{envelope}$ for a statistically significant number of clumps. To compare our predicted values with observational data, we distinguished between the PPP and PP case. For each model, we extracted cores that were found by their density and column density peaks and calculated $\overline{\Rb}$ and $\overline {|B_{LoS}|}$ for each of them. We also introduced a quantity $X(N)$, as a measure of the number of field reversals in the envelope of a clump and compared two different methods for computing $\R$.

We identify two primary physical processes by which $\R<1$ can be achieved in turbulent magnetised clouds: 
\begin{itemize}
\item In the strong magnetic field limit, field reversals are more likely in the envelope than in the core (see sec.~\ref{subsec:stronglimit}), such that magnetic field lines can cancel out there, leading to $\Rb\lesssim1$.
\item In the weak magnetic field limit, the small-scale dynamo amplification results in an increase of the initial magnetic field strength in the core, thus also leading to values $\Rb\lesssim1$ (see sec.~\ref{subsec:weaklimit}).
\end{itemize}

In addition, we report the following findings:
\begin{itemize}
\item We saw no significant time evolution in our distributions on the basis of $1\sigma$-interval for all $\beta_0$ and LoS-directions. This is because in all our simulations, turbulence was already fully developed and the magnetic field was saturated.
\item We did not measure any significant differences between PPP and PP, neither for any LoS nor for any $\beta_0$. We found that those values with $\Rb \lesssim 1$ are mainly generated by field reversals in the clumps, which lead to a higher average mass-to-flux ratio in the envelope than in the core.
\item Our distribution has average values of $\Rb\gtrsim 1$, but has a large standard deviation in $\Rb$ and $|B_{LoS}|$. \citet{Crutcher} observes average values of $\Rb\lesssim 1$, for four different cores/clumps. However, their four cores are consistent with our basic trend, but are located at the lower end of our distribution.
\item We find a similar trend as seen in the observations by \citet{Crutcher}, namely that $\Rb$ is smaller for smaller magnetic field strengths. As for the mean values of the observed $\Rb$, however, the trend does not necessarily reflect the average properties of typical clumps.
\item The actual values of $\Rb$ depend slightly on the analysis method. Our second method (sec.~\ref{subsec:computeR}) gives slightly larger values of $\Rb$ on average. Applying the second method to the four clumps observed in \citet{Crutcher}, we find that one of the four clumps has $\Rb\gtrsim1$, compared to the first method, for which all four clumps have $\Rb\lesssim1$.
\end{itemize}

For further investigations and to get a better understanding of the distribution of $\Rb$ in terms of the average magnetic field strength, we need additional efforts in both observations and simulations. While one has to observe many more clumps to get a better, statistically significant distribution, future numerical simulations must eventually incorporate turbulence and ambipolar diffusion, as well as an accurate treatment of the cores' chemical evolution together with a proper model for Zeeman splitting and radiative transfer.

\section*{Acknowledgements}
C.F.~received funding from the Australian Research Council (grant~DP110102191) and from the European Research Council (FP7/2007-2013 Grant Agreement no.~247060). C.F.,~R.B.,~and R.S.K.~acknowledge subsidies from the Baden-W\"urttemberg-Stiftung (grant P-LS-SPII/18) and from the German Bundesministerium f\"ur Bildung und Forschung via the ASTRONET project STAR FORMAT (grant 05A09VHA). R.S.K.\ further thanks for funding from the Deutsche Forschungsgemeinschaft via the SFB 881 The Milky Way System.
R.B.\ acknowledges funding by the Emmy-Noether grant (DFG) BA~3706.
Supercomputing time at the Leibniz Rechenzentrum (project no.~pr32lo) and the Forschungszentrum J\"ulich (project no.~hhd20) are gratefully acknowledged.
The FLASH code was in part developed by the DOE NNSA-ASC OASCR Flash Center at the University of Chicago.

\begin{appendix}
\section{Resolution study} \label{app:resol}

\begin{figure}
\centerline{
\includegraphics[width=1.0\linewidth]{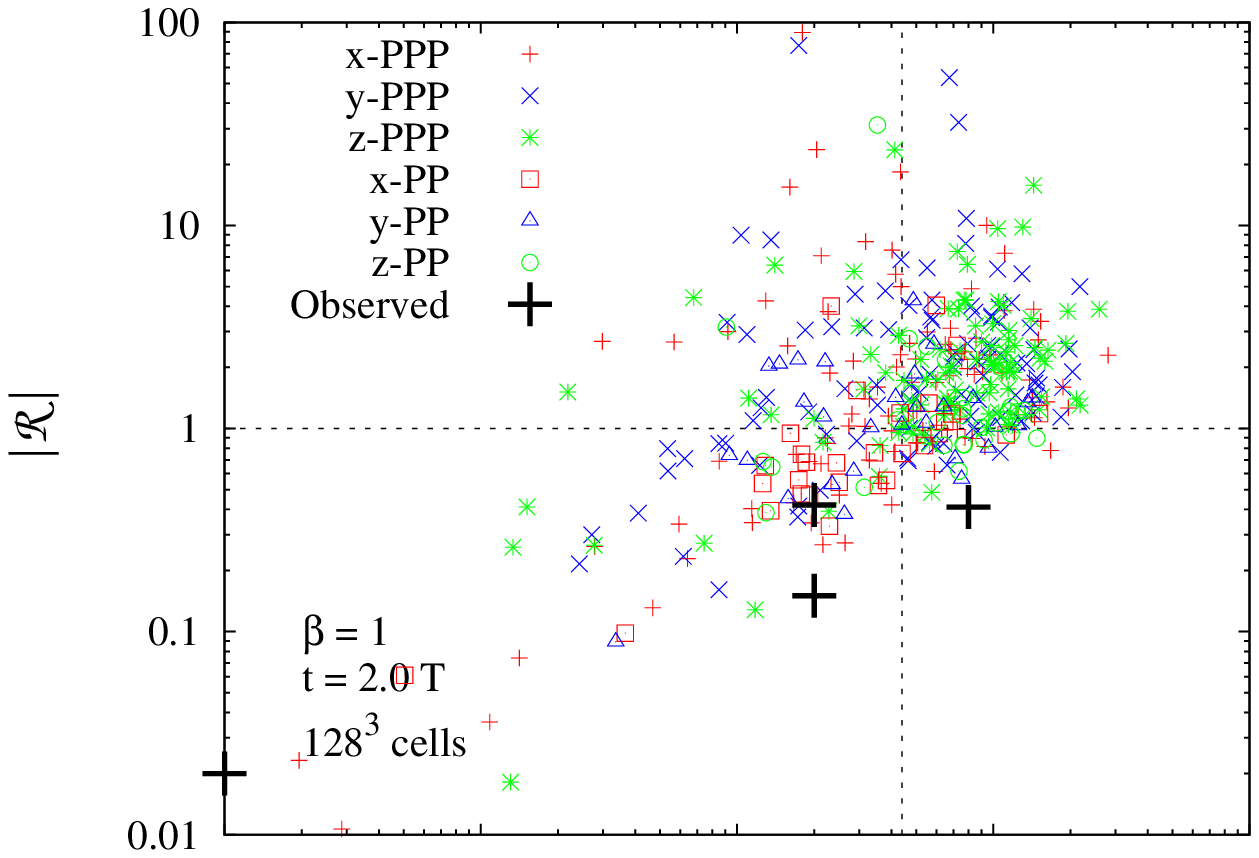} \\
}
\centerline{
\includegraphics[width=1.0\linewidth]{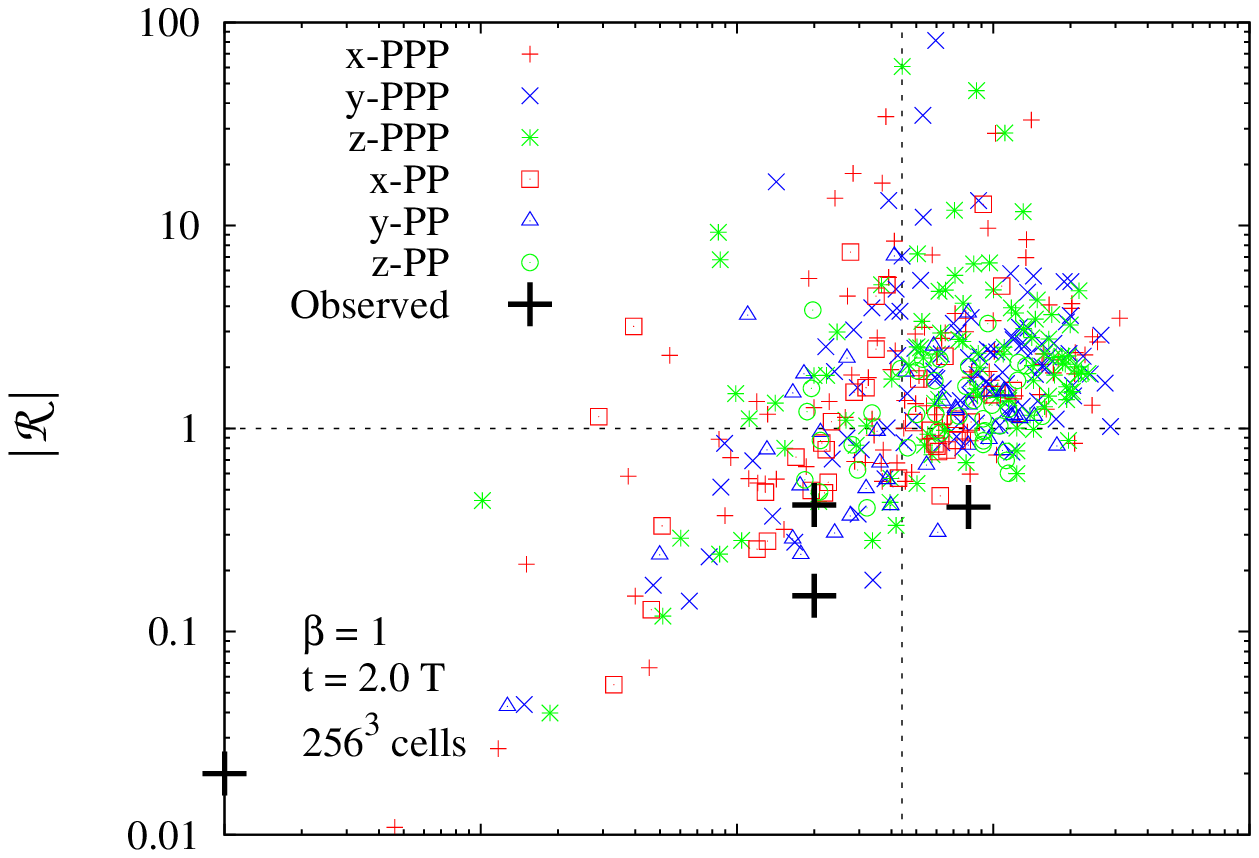} \\
}
\centerline{
\includegraphics[width=1.0\linewidth]{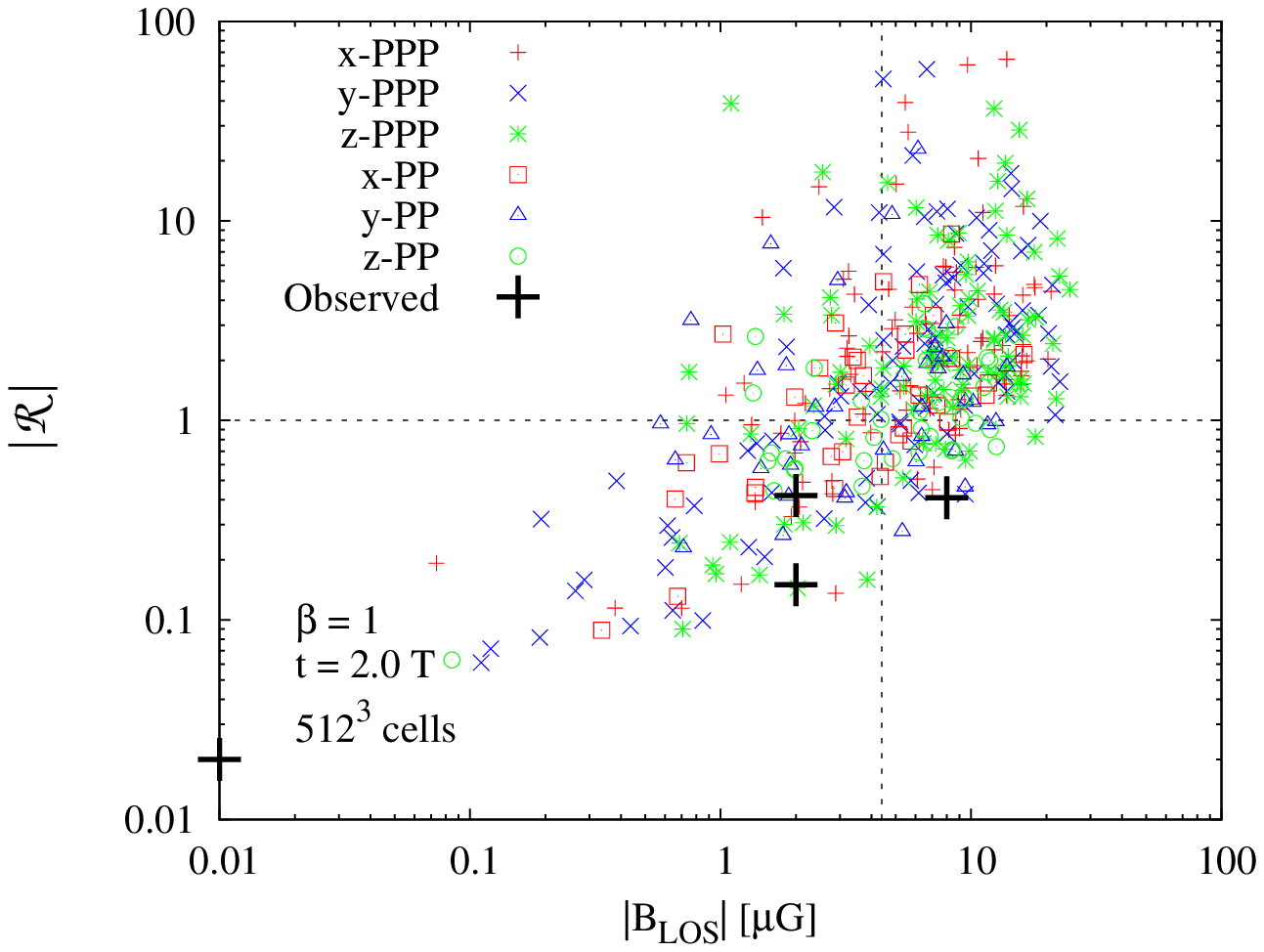} \\
}

\caption{Same as Figure~\ref{fig:scatter1}, but for a resolution study with $128^3$, $256^3$ and $512^3$ grid cells (from top to bottom) at $t=2.0\,T$. The mean, median, and standard deviations are listed in Table~\ref{tab:resolution}.}
\label{fig:resolution}
\end{figure}

In this section we present the results from our resolution studies (the values are computed with our first analysis method, sec.~\ref{subsec:computeR}). Therefore, as an example, we focused on the simulation with an initial plasma $\beta_0=1$ at $t = 2.0\,T$ for resolutions of $128^3$, $256^3$, and $512^3$ grid cells (the physical properties of our numerical experiment are not changed). Figure~\ref{fig:resolution} shows that there is no significant change in the distribution of our clumps, except the scattering due to statistical fluctuations. Table~\ref{tab:resolution} gives an overview of some statistical moments (mean, median and standard deviation) of the magnetic field component and $\Rb$ in each direction. The values do not differ significantly and are within the $1\sigma$-error-range, for example, $\overline {|B_{LoS}|}=8.1\pm5.2$, $9.5\pm6.5$, and $8.8\pm5.8 \mu$G for the different resolutions in z-direction. Also for the y-direction, where the variation is the strongest compared to the other LoS-directions, we have $\overline {|B_{LoS}|}=6.9\pm5.2$, $9.3\pm6.6$, and $7.4\pm5.8 \mu$G. Even in this case, the differences from the average values are smaller compared to the standard deviations of the distributions. Also the change of the average values of $B$ and $\Rb$ with time (as seen in section~\ref{sec:results}) is at most as big as the change caused by effects of resolution. Hence, our results do not depend significantly on resolution.

\begin{table}
\begin{tabular}{|l|l||l|l|l||l|l||l}
\hline\hline
 Res & LoS & $\overline {|B_{LoS}|}$ & $|\tilde{B}_{LoS}|$ & $\sigma_{|B|}$ & $\overline {\Rb}$ & $\tilde {\Rb}$ & $\sigma_{\Rb}$ \\
\hline
 & x & 5.8 & 4.3 & 5.1 & 3.3 & 1.4 & 9.4\\
 $128^3$ & y & 6.9 & 5.8 & 5.2 & 3.9 & 1.8 & 9.7\\
 & z & 8.1 & 7.4 & 5.2 & 2.6 & 1.9 & 3.1\\
\hline
 & x & 7.6 & 5.4 & 6.9 & 3.4 & 1.5 & 6.0\\
 $256^3$ & y & 9.3 & 8.1 & 6.6 & 3.7 & 1.9 & 9.0\\
 & z & 9.5 & 8.5 & 6.5 & 3.7 & 1.9 & 8.0\\
\hline
 & x & 7.6 & 6.9 & 5.2 & 4.8 & 2.0 & 10.0\\
 $512^3$ & y & 7.4 & 6.5 & 5.8 & 4.4 & 2.0 & 8.2\\
 & z & 8.8 & 7.9 & 5.8 & 4.2 & 1.9 & 6.7\\
\hline\hline
\end{tabular}
\caption{Mean, median and standard deviation for all directions for the magnetic field component and $\Rb$ for a resolution of $128^3$, $256^3$ and $512^3$ cells (from top to bottom separated by a line) for the PPP case ($t = 2.0\,T$) computed with our first method. All values of B are given in $\mu$G.}
\label{tab:resolution}
\end{table}

\bibliographystyle{apj}
\bibliography{lit/literature}

\end{appendix}
\end{document}